\documentclass[fleqn,usenatbib]{mnras}
\usepackage[T1]{fontenc}
\DeclareRobustCommand{\VAN}[3]{#2}
\let\VANthebibliography\thebibliography
\def\thebibliography{\DeclareRobustCommand{\VAN}[3]{##3}\VANthebibliography}

\usepackage{graphicx, float, lmodern, xcolor,anyfontsize, threeparttable, upgreek}
\usepackage{amsmath, amssymb, siunitx, tikz, mathtools,makecell}

\graphicspath{{figs/}}

\DeclareSIUnit\parsec{pc}
\DeclareSIUnit\angstrom{\text {Å}}
\newcommand{\ud}{\mathrm{d}}   
\newcommand{\ue}{\mathrm{e}}

\definecolor{lime}{HTML}{A6CE39}
\DeclareRobustCommand{\orcidicon}{\hspace{-3mm}
	\begin{tikzpicture}
	\draw[lime, fill=lime] (0,0) 
	circle [radius=0.16] 
	node[white] {\hspace{0.1mm}{\fontfamily{qag}\selectfont \tiny ID}};
	\draw[white, fill=white] (-0.07,0.1) 
	circle [radius=0.01];
	\end{tikzpicture}
	\hspace{-5mm}
}

\foreach \x in {A, ..., Z}{\expandafter\xdef\csname orcid\x\endcsname{\noexpand\href{https://orcid.org/\csname orcidauthor\x\endcsname}
		{\noexpand\orcidicon}}
}

\title[ECHO21]{ECHO21: A tool for modelling global 21-cm signal from dark ages to reionization}

\author[Mittal et al.]{
Shikhar Mittal$^{1,2}$\,\ \orcidA{}\, \thanks{E-mail: sm2941@cam.ac.uk}, Girish Kulkarni$^3$\,\ \orcidC{}\, , Peter Sims$^{1,2}$\,\ \orcidE{}
\\
\\
$^{1}$Battcock Centre for Experimental Astrophysics, Cavendish Laboratory, J.~J.\ Thomson Avenue, Cambridge CB3 0HE, UK\\
$^{2}$Kavli Institute for Cosmology, University of Cambridge, Madingley Road, Cambridge CB3 0HA, UK\\
$^3$Tata Institute of Fundamental Research, Homi Bhabha Road, Mumbai 400005, India
}

\date{Accepted 2026 January 02. Received 2025 December 01; in original form 2025 June 28}

\pubyear{2025}

\begin{document}
\label{firstpage}
\pagerange{\pageref{firstpage}--\pageref{lastpage}}
\maketitle

\begin{abstract}
We introduce a Python package called \texttt{ECHO21} for modelling the global 21-cm signal from the dark ages through cosmic dawn to the end of reionization. Leveraging its analytical framework, \texttt{ECHO21} generates a single model in $\mathcal{O}(1)\,$s, allowing a large number of signals to be generated efficiently by distributing models across multiple cores. Thus, it is ideal for performing astrophysical or cosmological inference from a given 21-cm dataset. We offer six astrophysical parameters that control the Lyman-$\upalpha$ (Ly$\upalpha$) emissivity, X-ray emissivity, emissivity of ionizing photons, and star formation rate. Beyond its efficiency some of the attractive and novel features in \texttt{ECHO21} relative to previously published codes are inclusion of Ly$\upalpha$ heating, ability to vary the standard cosmological parameters as easily as the astrophysical parameters, and different models of star formation rate density (physically-motivated, a semi-empirical, and an empirically-motivated). With a number of 21-cm experiments soon to provide cosmic dawn 21-cm data, \texttt{ECHO21} is a flexible and extensible new open-source package for making quick but sufficiently realistic astrophysical inferences. We make our code publicly available.
\end{abstract}



\section{Introduction}
Dark ages and the cosmic dawn are not well understood periods of cosmic timeline because of the lack of direct observational probes. One probe that could potentially unravel the physics of the first stars is the hyperfine transition line of wavelength $\SI{21}{\centi\metre}$ originating in a neutral hydrogen atom. The notion of 21-cm line as a cosmological probe originated in the 1990s \citep{MMR_1997} and has ever since seen significant improvements in its modelling. See review articles by \citet{Furlanetto}, \citet{Pritchard_2012}, \citet{Barkana_2016} and \citet{Mesinger_19}.

Besides its usefulness in understanding the ionization and thermal history, the 21-cm line has also been proposed to study primordial magnetic fields \citep[e.g.][]{Bhaumik_2025}, gravitational waves \citep[e.g.][]{Mishra_2018}, and the fundamental nature of dark matter \citep[e.g.][]{Katz_2024}, among others. The wide applicability of the 21-cm line as a cosmological probe has led to the development of many experiments targeting the global 21-cm signal, such as Experiment to Detect the Global EoR Signal \citep[\textit{EDGES},][]{Bowman}, Shaped Antenna measurement of the background RAdio Spectrum \citep[\textit{SARAS},][]{saras3}, Radio Experiment for the Analysis of Cosmic Hydrogen \citep[\textit{REACH},][]{reach}, Large Aperture Experiment to Detect the Dark Ages \citep[\textit{LEDA},][]{Price_2018}, Probing Radio Intensity at high-Z from Marion \citep[\textit{PRIZM},][]{Philip_2019}, the Mapper of the IGM Spin Temperature \citep[\textit{MIST},][]{Monsalve_2024}, and the Remote \ion{H}{i} eNvironment Observer \citep[\textit{RHINO},][]{Bull_2025}.

One of the primary goals of the current cosmology is to analyse and robustly interpret 21-cm signal data from experiments and make inferences about standard or even exotic astrophysical and/or cosmological processes. This has motivated the development of public codes which predict global cosmic dawn 21-cm signal for a wide range and choice of astrophysical parameters. Some of the notable numerical or semi-numerical codes include \texttt{21cmFAST} \citep{Mesinger_2011} and \texttt{BEORN} \citep{Schaeffer_2023}. Numerical codes while accurate and physically rich, can be slow to run. This necessitates the requirement of analytic, sufficiently realistic models, and fast enough for quick and reliable inferences. Existing codes of this kind that have an analytical framework include \texttt{ARES} \citep{Mirocha_2014} and \texttt{Zeus21} \citep{Munoz_2023}.

In this work we introduce a Python package called \texttt{ECHO21}, which stands for Exploring the Cosmos with Hydrogen Observation, which is a code having an analytic framework. \texttt{ECHO21} can produce a single instance of the global 21-cm signal in the redshift range $0\leqslant z\leqslant1500$ in a wall-clock time of $\sim \mathcal{O}(1)\,$s on a single CPU. 

A major difference between \texttt{ECHO21} and \texttt{ARES} or \texttt{Zeus21} is in the heating due to Lyman-$\upalpha$ (Ly$\upalpha$) photons. At the time of writing, neither \texttt{ARES} nor \texttt{Zeus21} account for the Ly$\upalpha$ heating. As we show later, non-inclusion of Ly$\upalpha$ heating can lead to strong biases in the signal modelling. We also identify, for the first time, the different approaches adopted for modelling the star formation rate density and how they impact the global signal.
 
A secondary goal behind the development of this code is to provide a physically-motivated global 21-cm signal for the \textit{REACH} data analysis pipeline testing. While an empirical Gaussian model \citep[e.g.,][]{Mittal_eps} can be employed for a quick test of the pipeline, a physically-motivated model of the signal will allow us to make realistic astrophysical inferences with 21-cm observational data. Further, while \texttt{ECHO21} models may be less accurate than semi-numerical models, \texttt{ECHO21} is easily modified to include new physics.

This paper is organized as follows. In section~\ref{sec:theory} we go over the modelling details of the global 21-cm signal from the dark ages to the end of epoch of reionization (EoR). In section~\ref{sec:res} we present the main outputs of \texttt{ECHO21}. We conclude in section~\ref{sec:conc}.

We adopt a flat-$\Lambda$CDM cosmology with the following parameters: $h=0.674\pm0.005$, $\Omega_{\mathrm{m}}= 0.315\pm0.007$, $\Omega_{\mathrm{b}}=0.049\pm0.001$, $\sigma_8 = 0.811\pm0.006$, and $n_{\mathrm{s}} = 0.965\pm0.004$ \citep{Planck}. In addition to these we have the CMB temperature today $T_{\gamma0}=2.7255\pm0.0006\,$K \citep{Fixsen_2009} and the primordial helium abundance $Y_{\mathrm{p}}=0.245\pm0.004$ \citep{Aver_2015}. All values are quoted with 68\% confidence limits. Unless stated otherwise we work with the median values of the parameters. We prefix the distance units with a `c' to indicate a comoving length while no prefix to indicate proper physical lengths.

\section{Theory and methods}\label{sec:theory}
We begin by writing down the observable 21-cm signal, which is the 21-cm brightness temperature measured against the cosmic microwave background (CMB) temperature and is given by \citep{F06}
\begin{multline}
T_{21}=27\bar{x}_{\ion{H}{i}}\left(\frac{1-Y_{\mathrm{p}}}{0.76}\right)\left(\frac{\Omega_\mathrm{b}h^2}{0.023}\right)\sqrt{\frac{0.15}{\Omega_\mathrm{m}h^2}\frac{1+z}{10}}\\\times\left(1-\frac{T_{\mathrm{r}}}{T_\mathrm{s}}\right)\si{\milli\kelvin}\,,\label{eq:DeltaT}
\end{multline} 
where $T_\mathrm{s}$ is the spin temperature, $T_{\mathrm{r}}=T_\gamma$ is CMB temperature, $z$ is the redshift, and $x_{\ion{H}{i}}\equiv n_{\ion{H}{i}}/n_{\text{H}}$ is the ratio of number densities of neutral hydrogen (\ion{H}{i}) and total hydrogen (H). In standard calculations the background is taken to be the CMB, hence $T_{\mathrm{r}}=T_\gamma$, but for an additional radio flux $T_{\mathrm{r}}$ will contain another term \citep{Feng_2018, Ewall, Fialkov_19, Mittal_erb, Singal_2023}. We do not consider an excess radio term in this work.

We assume a two-zone model for the intergalactic medium (IGM). In such a scenario the IGM is considered a composition of the bulk IGM and an \ion{H}{ii} region. \ion{H}{ii} regions refer to the pockets of fully ionized medium around the galaxies because of the ionizing-UV photons possibly emitted by star-forming galaxies (or active galactic nuclei). Everything outside these \ion{H}{ii} regions is the bulk IGM. If the electron fraction of the bulk IGM is $x_{\mathrm{e}}$ and the fraction of volume of \ion{H}{ii} region with respect to the total volume is $Q$, then a `globally-averaged' ionized fraction of the Universe can be written as \citep{Mirocha_2013,Mirocha_2015}
\begin{equation}
\bar{x}_{\mathrm{i}}=Q+(1-Q)x_{\mathrm{e}}\,.\label{eq:gaif}
\end{equation}
Accordingly, the globally-averaged neutral hydrogen fraction is $\bar{x}_{\ion{H}{i}}=1-\bar{x}_{\mathrm{i}}$, which appears in equation~\eqref{eq:DeltaT}. 

The spin temperature is \citep{Mittal_lya}
\begin{equation}
T_\mathrm{s}^{-1}=\frac{T_\gamma^{-1}+x_\mathrm{k}T_\mathrm{k}^{-1}+x_{\mathrm{Ly}}(T_\mathrm{k}+T_{\text{se}})^{-1}}{1+x_\mathrm{k}+x_{\mathrm{Ly}} T_\mathrm{k}(T_\mathrm{k}+T_{\text{se}})^{-1}}\,,
\end{equation}
where $T_{\text{k}}$ is the gas temperature, $T_{\text{se}}\approx\SI{0.4}{\kelvin}$ accounts for the spin-exchange correction \citep{Chuzhoy}, $x_\mathrm{k}$ is the collisional coupling, and $x_{\mathrm{Ly}}$ is the Ly$\upalpha$ coupling \citep{Wouth, Field}.

The collisional coupling \citep{Zygelman_2005} can be expressed as
\begin{equation}
x_{\mathrm{k}}=\frac{T_\star n_\mathrm{H}}{T_\gamma A_{10}}\left[(1-x_{\mathrm{e}})\kappa_{\mathrm{HH}}+x_{\mathrm{e}}\kappa_{\mathrm{eH}}+x_{\mathrm{e}}\kappa_{\mathrm{pH}}\right]\,,\label{XK}
\end{equation}
where $T_\star=\SI{0.068}{\kelvin}$, $A_{10}=\SI{2.85e-15}{\second^{-1}}$ is the Einstein coefficient of spontaneous emission for the hyperfine transition, $n_\mathrm{H}$ is the proper hydrogen number density, and $\kappa_{i\mathrm{H}}$'s are the de-excitation rate of hyperfine transition because of collision between $i$ and \textsc{H\,i} \citep{FF06, Furlanetto, FF07}.

The Ly$\upalpha$ coupling is 
\begin{equation}
x_{\mathrm{Ly}} = S\frac{J_{\mathrm{Ly}}(z)}{5.54\times10^{-8}(1+z)\,\si{\metre^{-2}\second^{-1}\hertz^{-1}\steradian^{-1}}}\,,
\end{equation}
where $S$ is the scattering correction and $J_{\mathrm{Ly}}$ is the total Ly$\upalpha$ background. For scattering correction we adopt the approximate form by \citet{Chuzhoy}, so that
\begin{equation}
S\approx\ue^{-1.69\zeta^{2/3}}\,,
\end{equation}
where
\begin{equation}
\zeta=\frac{4}{3}\sqrt{\frac{a_{\mathrm{V}}\tau\eta^3}{\pi}}\,.
\end{equation}
The Voigt parameter, optical depth, and recoil parameter for Ly$\upalpha$ line are $a_{\mathrm{V}},\tau$ and $\eta$. For more details see \citet{Mittal_lya} or \citet{Mittal_pbh}.

The Ly$\upalpha$ coupling recipe described above is based upon the wing approximation of Ly$\upalpha$ radiative transfer. In this approximation the photons are assumed to be far from the resonance -- the wings, where the photons have long mean free paths. However, in our previous work \citep{Mittal_mcrt}, we showed that multiple line core scatterings affect the intensity profile of the Ly$\upalpha$ photons causing a reduction in the Ly$\upalpha$ coupling. See also \citet{Reis_2021} and \citet{Semelin2023}. Incorporating such a numerical model into \texttt{ECHO21}, which has an analytical framework, is beyond the scope of this work and thus, we leave it for a future version of \texttt{ECHO21}.

For the construction of Ly$\upalpha$ specific intensity, $J_{\mathrm{Ly}}$, we follow \citet{Mittal_lya}; we assume that the specific intensity tracks the instantaneous star formation rate with the stellar population having a power-law spectral energy distribution (SED). We account for the contribution of higher Lyman-series photons (up to $n=23$) in the computation of $J_{\mathrm{Ly}}=J_{\mathrm{Ly}}(z)$ via radiative cascading \citep{PF06}. We assume a uniform spectral index between Ly$\upalpha$ and Lyman limit lines. Thus, if $N_{\mathrm{Ly}}$ is the total number of Lyman series photons and the spectral index is $-s-1$ then the SED is
\begin{equation}
\phi_{\mathrm{Ly}}(E)\propto f_{\mathrm{Ly}}\frac{N_{\mathrm{Ly}}}{E_\infty}\left(\frac{E}{E_{\infty}}\right)^{-s-1}\,,\label{eq:lysed}
\end{equation}
when expressed as number of photons emitted per unit energy range per stellar baryon. The normalization factor for the above is\footnote{If $s=0$, then the normalization factor in equation~\eqref{eq:lysed} is $1/\ln(E_\infty/E_\upalpha)\,.$},
\begin{equation}
\frac{s}{(E_\infty/E_\upalpha)^s-1}\,,
\end{equation}
where $E_\infty=\SI{13.6}{\electronvolt}$ and $E_\upalpha=\SI{10.2}{\electronvolt}$ are the Lyman limit and Ly$\upalpha$ energy, respectively. For our fiducial model we have $N_{\mathrm{Ly}}=10000$ and $s=2.64$. This is inspired by the Pop-II stellar model \citep{BL05}, where we have defined a single `average' power law index for the broken power-law model. In the code we do not consider $N_{\mathrm{Ly}}$ to be a free parameter but instead allow it to scale by a dimensionless factor $f_{\mathrm{Ly}}$. This introduces the first free parameter in our model, fiducial choice for which is 1. The other free parameter associated with Ly$\upalpha$ specific intensity is $s$.

We have used a simple two-parameter model for the Lyman series SED. However, in general the SED is a complex function of stellar properties such as the initial mass function, metallicity, and age. This dependence could be accounted for by allowing \texttt{ECHO21} to interface with stellar population synthesis codes or data such as \texttt{SPISEA} \citep{Hosek_2020} or \texttt{BPASS} \citep{eldridge}.\\
[1\baselineskip]

The global 21-cm signal requires the knowledge of ionization history and `spin history'. Spin history in turn is dependent on thermal history and the knowledge of coupling terms. Our task then boils down to the computation of $T_{\mathrm{k}}, x_{\mathrm{e}}$, and $Q$, which we discuss next.

\subsection{Thermal and ionization history}\label{sec:igm_equations}
In this section we discuss the differential equations governing the evolution of $T_{\mathrm{k}}, x_{\mathrm{e}}$, and $Q$ of our two-zone IGM model.

\subsubsection*{Gas kinetic temperature}
The equation describing the evolution of gas temperature $T_{\mathrm{k}}$ of the bulk IGM is
\begin{equation}
\frac{\ud T_{\mathrm{k}}}{\ud t}=-2HT_{\mathrm{k}}-\frac{T_{\mathrm{k}}}{1+x_{\text{He}}+x_{\text{e}}}\frac{\ud x_{\text{e}}}{\ud t}+\frac{2}{3n_{\mathrm{b}}k_{\mathrm{B}}}\sum q\,,\label{eq:tk}
\end{equation}
where $k_{\mathrm{B}}$ is the Boltzmann constant, $x_{\text{He}}=Y_{\mathrm{p}}/[4(1-Y_{\mathrm{p}})]$ is the ratio of helium to hydrogen number densities, $n_{\mathrm{b}}$ is the total baryon number density, and $q$ is the volumetric heating rate. The first term on the right hand side describes the adiabatic cooling, second term corresponds to the change in energy distribution due to changing particle number and last term accounts for all other heating or cooling processes. In this work we consider Compton heating, Ly$\upalpha$ heating and X-ray heating. 

The Compton heating term is \citep{Seager_2000}
\begin{equation}
q_{\mathrm{Comp}}=16\frac{\sigma_\mathrm{T}\sigma_\mathrm{S} T_\gamma^4}{m_{\mathrm{e}}c^2}n_{\mathrm{H}}x_{\mathrm{e}}k_{\mathrm{B}}(T_\gamma-T_\mathrm{k})\,,
\end{equation}
where $\sigma_\mathrm{T}$, $\sigma_\mathrm{S}$, $m_{\mathrm{e}}$, and $c$ are Thomson scattering cross-section, Stephan-Boltzmann constant, mass of electron, and speed of light, respectively.

The volumetric (proper) heating rate due to Ly$\upalpha$ photons is \citep{Mittal_lya}
\begin{equation}
q_{\mathrm{Ly}}=4\pi h_{\mathrm{P}}\Delta\nu_\mathrm{D} \frac{H}{\lambda_\upalpha} \left(J_{\mathrm{Ly}}^\mathrm{c}I_\mathrm{c}+J_{\mathrm{Ly}}^\mathrm{i} I_\mathrm{i}\right)\,,\label{eq:lyaheat}
\end{equation}
where $h_{\mathrm{P}}, \Delta\nu_\mathrm{D}$, and $\lambda_\upalpha$ are Planck's constant, Doppler width for Ly$\upalpha$ line, and the Ly$\upalpha$ wavelength, respectively. The Ly$\upalpha$ intensity due to continuum and injected photons is $J_{\mathrm{Ly}}^\mathrm{c}$ and $J_{\mathrm{Ly}}^\mathrm{i}$, respectively, such that $J_{\mathrm{Ly}}=J_{\mathrm{Ly}}^\mathrm{c}+J_{\mathrm{Ly}}^\mathrm{i}$. $I_{\mathrm{c}}$ is the normalized area between the undisturbed continuum Ly$\upalpha$ intensity and a scattered one. Similarly, $I_{\mathrm{i}}$ is defined for injected photons. We use the wing approximation for the computation of $I_{\mathrm{c}}$ and $I_{\mathrm{i}}$.

Finally, we also include heating due to X-ray photons. The volumetric (comoving) X-ray heating rate is
\begin{equation}
q_{\mathrm{X}} = f_{\mathrm{X}} \cdot f_{\mathrm{X,h}}(x_{\ue})\cdot C_{\mathrm{X}}(E_0,E_1) \cdot \dot{\rho}_\star(z)\,,\label{eq:xheat}
\end{equation}
where
\begin{equation}
f_{\mathrm{X,h}}=1-\left(1-x_{\ue}^{0.2663}\right)^{1.3163}\,,
\end{equation}
is the fraction of integrated X-ray emissivity resulting in IGM heating \citep{Shull_1985}. As evident from the above expression, we follow a simple parametric approach to X-ray heating \citep{F06}; we assume that the integrated X-ray emissivity (energy per unit time per unit comoving volume) tracks the instantaneous star formation rate density (just as in the case of Ly$\upalpha$ intensity) so that $\epsilon_{\mathrm{X}}\propto\dot{\rho}_\star$. We calibrate this correlation to local galaxy observations, which indicate that galaxy-integrated X-ray luminosity ($L_{\mathrm{X}}$) scales linearly with star formation rate (SFR) \citep[see, e.g.,][]{Mineo}. We adopt a recent value based on the \textit{Chandra} observations of 88 nearby galaxies. For the energy range $\tilde{E}_0=0.5$ to $\tilde{E}_1=\SI{8}{\kilo\electronvolt}$ \citep{Lehmer_2024}
\begin{equation}
C_{\mathrm{X}}(\tilde{E}_0,\tilde{E}_1)=\frac{L_{\mathrm{X}}}{\mathrm{SFR}} = 2.45\times10^{39}\,\mathrm{erg\,s^{-1}\left(M_{\odot}yr^{-1}\right)^{-1}}\,.\label{eq:lxsfr}
\end{equation}
To account for the contribution of low-energy photons (which have higher interaction cross-section) and high-energy photons (which have longer mean free path) we extrapolate $C_{\mathrm{X}}$ for energies ranging from $E_0=0.2$ to $E_1=\SI{30}{\kilo\electronvolt}$ \citep{Mirocha_19}. Assuming a spectral index of $-w$, so that the specific luminosity is given by $l_{\mathrm{X}}(E)\propto E^{-w}$, we get\footnote{If $w=1$, then equation~\eqref{eq:xsed} becomes$$C_{\mathrm{X}}(E_0,E_1)= C_{\mathrm{X}}(\tilde{E}_0,\tilde{E}_1)\frac{\ln(E_1/E_0)}{\ln(\tilde{E}_1/\tilde{E}_0)}\,.$$}, 
\begin{equation}
C_{\mathrm{X}}(E_0,E_1)= C_{\mathrm{X}}(\tilde{E}_0,\tilde{E}_1)\frac{E_1^{1-w}-E_0^{1-w}}{\tilde{E}_1^{1-w}-\tilde{E}_0^{1-w}}\,.\label{eq:xsed}
\end{equation}
For our fiducial model we set $w=1.5$ which gives $C_{\mathrm{X}}(E_0,E_1)=4.74\times10^{39}\,\mathrm{erg\,s^{-1}\left(M_{\odot}yr^{-1}\right)^{-1}}$. Finally, we have another free parameter $f_{\mathrm{X}}$ whose fiducial value is $1$.

Note that the $C_{\mathrm{X}}$ value we adopt is calibrated to low redshifts. At cosmic dawn the stars are metal-poor \citep{Bromm_2013} and large-scale population synthesis simulations have shown that $C_\mathrm{X}$ increases with decreasing metallicity \citep{Fragos_2013}. However, high-redshift uncertainties in $C_{\mathrm{X}}$ can be subsumed in the factor $f_{\mathrm{X}}$.

In addition to the above we note the following point. A physically-motivated approach and hence a more accurate estimate of X-ray heating involves calculating background specific intensity of X-rays, $J_{\mathrm{X}}$. It is well-known that computing $J_{\mathrm{X}}$ throughout the cosmic dawn can slow down the simulation by a significant amount \citep{Mesinger_2013}. An empirical method like ours works quite well for steeper spectra but fails for flatter ones; $\sim\SI{5}{\milli\kelvin}$ difference is seen in the peak global 21-cm signal for $w=1.5$ \citep{Mirocha_2014}. However, given the high level of uncertainty in the spectra of cosmic dawn Universe, here we subsume these differences within the variable $f_{\mathrm{X}}$; we do not expect this to be the dominant source of differences.

\subsubsection*{Bulk IGM electron fraction}
Having discussed the thermal evolution, we now move to the ionization history of the bulk IGM. The equation governing the bulk IGM electron fraction $x_{\ue}$ is given by the standard recombination equation \citep{Cyr_2024}. In addition to the standard photoionization and recombination, we also consider the ionization by the X-ray photons released by star-forming galaxies, which becomes important after cosmic dawn \citep{Mirabel}. Thus, our complete equation governing the bulk IGM electron fraction is given by
\begin{equation}
\frac{\ud x_{\text{e}}}{\ud t}=C_{\mathrm{P}}\left[\beta(1-x_{\text{e}})\ue^{-E_\upalpha/(k_\mathrm{B}T_{\gamma})}-x_{\text{e}}^2n_{\mathrm{H}}\alpha_{\mathrm{B}}\right]+\Gamma_{\mathrm{X}}(1-x_{\text{e}})\,,\label{eq:xe}
\end{equation}
where $C_{\mathrm{P}}$ is the Peebles' factor defined as
\begin{equation}
C_{\mathrm{P}}=\frac{1+K\Lambda_{\mathrm{2s-1s}}n_{\mathrm{H}}(1 - x_{\mathrm{e}})}{ 1+K\Lambda_{\mathrm{2s-1s}}n_{\mathrm{H}}(1 - x_{\mathrm{e}}) + K\beta n_{\mathrm{H}}(1 - x_{\mathrm{e}})}\,,
\end{equation}
$\beta$ is the photoionization rate, and $\Gamma_{\mathrm{X}}$ is the ionization due to X-ray photons. In Peebles' factor $\Lambda_{\mathrm{2s-1s}}=\SI{8.22458}{\second^{-1}}$ is the hydrogen 2s–1s two-photon
rate and $K=\lambda_\upalpha^3/8\pi H(z)$ is the cosmological redshifting of Ly$\upalpha$ photons, where $\lambda_\upalpha=\SI{121.5682}{\nano\metre}$ is the wavelength of Ly$\upalpha$ photon. We follow \citet{Pequignot_1991} for the temperature dependence of case-B recombination coefficient $\alpha_{\mathrm{B}}$, with an additional multiplicative correction factor of $F=1.14$ to capture the effective-three-level-atom recombination model \citep{Seager_1999}.

We evaluate the photoionization rate (also appearing in the Peebles' factor) at the CMB temperature and \textit{not} the gas temperature; the recombination coefficient is still computed at the gas temperature \citep{Chluba_2015}. This modification in equation~(71) \citet{Seager_2000} captures the leading order correction in the full physics of the recombination in the ambient CMB bath \citep{Haimoud_2010, Chluba_2012}.

We estimate the total ionization rate due to X-ray photons as (see App.~\ref{app:xray_ion})
\begin{equation}
\Gamma_{\mathrm{X}}\simeq\left(\frac{1}{E_w}+\frac{f_{\mathrm{X,ion}}}{E_{\infty}}\right)\frac{q_{\mathrm{X}}}{f_{\mathrm{X,h}}n_{\ion{H}{i}}}\,,
\end{equation}
where\footnote{Just as the case for Lyman series or X-ray SED, one can define an appropriate limit of $E_w$ for $w=-3.4$ or $-2.4$.}
\begin{equation}
E_w=\frac{w+3.4}{w+2.4}\left(\frac{E_1^{-w-2.4}-E_0^{-w-2.4}}{E_1^{-w-3.4}-E_0^{-w-3.4}}\right)-E_{\infty}\,,\label{eq:Ew}
\end{equation}
and the enhancement factor
\begin{equation}
f_{\mathrm{X,ion}}=0.3908\left(1-x_{\mathrm{e}}^{0.4092}\right)^{1.7592}\,.
\end{equation}
accounts for the secondary ionization \citep{Shull_1985, Madau_2017}.

\subsubsection*{Volume-filling factor: modelling reionization}
To model reionization we adopt the analytic approach which does a comparison of ionizing photons against the radiative recombinations. Accordingly the reionization equation describing the volume-filling factor ($Q$) of \ion{H}{ii} region obtained is \citep{Madau_1999}
\begin{equation}
\frac{\ud Q}{\ud t}=\frac{\dot{n}_{\mathrm{ion}}}{n^0_{\mathrm{H}}}-A_{\mathrm{He}}n_{\mathrm{H}}\alpha_{\mathrm{B},10^4}CQ\,,\label{eq:reion}
\end{equation}
where $\dot{n}_{\mathrm{ion}}$ is the comoving volume-averaged emission rate of ionizing photons into the IGM, and $n^0_{\mathrm{H}}$ is the number density of hydrogen today. We assume that the \ion{H}{ii} region is maintained at a fixed temperature of $\SI{e4}{\kelvin}$ so that $\alpha_{\mathrm{B},10^4}=\SI{2.94e-19}{\metre^3\second^{-1}}$. The clumping factor, to correct for the enhanced recombination rate, of ionized hydrogen is \citep{Shull_2012}
\begin{equation}
C(z)=20.81(1+z)^{-1.1}\,.
\end{equation}
We assume that the reionization of \ion{H}{i} and \ion{He}{i} proceed simultaneously so that these species can be described by a single equation. The factor $A_{\mathrm{He}}=1+x_{\mathrm{He}}$ in the recombination term accounts for the electron contribution from the first ionization of helium. In this work we do not account for its second ionization due to its high ionization energy resulting in a delayed \ion{He}{ii} reionization \citep{Haard}. However, we assume an instantaneous \ion{He}{ii} reionization at $z=4$ for the computation of electron-scattering optical depth (see below).

The source of ionizing photons is an ongoing debate \citep{robertson}. It is generally advocated that the galaxies are the main drivers of reionization \citep[e.g.][]{Robertson_2015, Finkelstein_2016, Naidu_2020, Dayal_2020, Yeh_2023, Atek_2024, Dayal_2024}. However, the recent \textit{JWST} observations potentially hint towards AGN-driven reionization \citep{Munoz_2024, Madau_2024, Asthana_2024}.

We assume a galaxy-dominated ionizing emissivity and that the emissivity tracks the instantaneous star formation rate density (SFRD), so that
\begin{equation}
\dot{n}_{\mathrm{ion}}=f_{\mathrm{esc}}I_{\mathrm{ion}}\dot{\rho}_\star\,,
\end{equation}
where the ionizing photon yield, $I_{\mathrm{ion}}$, is the rate at which ionizing photon are emitted from star-forming galaxies per unit star formation rate and $f_{\mathrm{esc}}$ is the fraction of ionizing photons that actually escape the galaxies leaking into the IGM. $f_{\mathrm{esc}}$ becomes our next free parameter whose fiducial value we set to be 0.01. We have chosen this value such that we have a `late' reionization scenario \citep{Kulkarni}, which is consistent with the current observational evidence \citep[e.g.][and references therein]{Sims_2025}; our reionization is complete by $z\approx5.3$ (see Fig.~\ref{fig:ion}). In general, $f_{\mathrm{esc}}$ is expected to be a function of halo mass and/or redshift \citep[see, e.g.,][]{Dayal_2020, Saldana_2023, Lin_2023, Mutch_2023}. In the first version of \texttt{ECHO21} we consider a constant value of $f_{\mathrm{esc}}$ and leave more general models for a future version.  

The ionizing photon yield in general is determined by stellar physics and initial mass function (IMF). For the Kroupa IMF and a metallicity of $0.01Z_{\odot}$ we have \citep{Madau_2017}, 
\begin{equation}
I_{\mathrm{ion}}=10^{53.44}\,\mathrm{s}^{-1}\left(\mathrm{M}_{\odot}\mathrm{yr}^{-1}\right)^{-1}\,.
\end{equation}
This is equivalent to $N_{\mathrm{ion}}=7250$ ionizing photons for every hydrogen atom that collapses to form star particle.

The CMB observations provide a complementary probe to the epoch of reionization. For a given reionization history we compute the electron-scattering optical depth to $z$ as \citep{Kuhlen_2012}
\begin{equation}
\tau_\ue(z)=c\sigma_{\mathrm{T}}n^0_{\mathrm{H}}\int_0^{z}\left[1+N(z')x_{\mathrm{He}}\right]Q(z')\frac{(1+z')^2}{H(z')}\ud z'\,,\label{eq:tau}
\end{equation}
where $N(z)=1+\Theta(4-z)$, for Heaviside step function $\Theta$. Function $N$ accounts for an additional electron contribution from singly ionized helium for $z\geqslant4$ and for two electrons from doubly ionized helium for $z<4$.\\
[1\baselineskip]

We now have three ordinary differential equations, \eqref{eq:tk}, \eqref{eq:xe}, and \eqref{eq:reion}. Note that the two zones, bulk IGM and \ion{H}{ii} regions, evolve independently of each other. As a result, only two of the equations, namely \eqref{eq:tk} and \eqref{eq:xe} are coupled, while \eqref{eq:reion} can be independently solved. We initialize the system at a high redshift, $z=1500$, when the gas and CMB temperature are equal, i.e. $T_{\mathrm{k}}=T_{\gamma}=\SI{4090.23}{\kelvin}$. For a high-density medium such as that at $z=1500$, we expect Saha's equation (equilibrium version of equation~\ref{eq:xe}) to be valid. For our fiducial choice of cosmological parameters we get $x_{\ue}=0.949$. Note that we activate Ly$\upalpha$ heating, X-ray heating and X-ray ionization only after the beginning of star formation, which we choose to begin at $z=z_{\star}=59$. (We choose a large value of $z_{\star}$ in order to ensure a smooth transition from dark ages to cosmic dawn.) Thus, for $z>z_{\star}$ SFR is 0 and so are $q_{\mathrm{Ly}}, q_{\mathrm{X}}$ and $\Gamma_{\mathrm{X}}$. Similarly, reionization has no meaning before cosmic dawn, and thus $\dot{Q}=0$ for $z>z_{\star}$. Additionally, note that equations \eqref{eq:tk}, \eqref{eq:xe}, and \eqref{eq:reion} are valid only until $Q$ reaches 1. When $Q=1$, we set $\dot{Q}=0$.

\subsection{Star formation rate density}\label{sec:sfrd}
In our model we assume that all three emissivities, namely Lyman-series, X-ray, and ionization track the instantaneous star formation rate density (SFRD). In this section go over our star formation prescription as it directly controls the IGM heating and ionization. In \texttt{ECHO21} we consider three types of model for SFRD, namely a physically-motivated, a semi-empirical, and an empirically-motivated model.

\subsubsection*{Physically-motivated model}
For our physically-motivated model we estimate the SFRD as the rate at which baryons collapse into the dark matter haloes and convert to star particles with a star formation efficiency (SFE), $f_{\star}$. Thus, a physically-motivated (comoving) SFRD is
\begin{equation}
\dot{\rho}_\star(z)=\frac{\Omega_\mathrm{b}}{\Omega_{\mathrm{m}}}\frac{\ud}{\ud t}\int_{M_{\mathrm{h,min}}}^{\infty} f_{\star}M_{\mathrm{h}}\frac{\ud n}{\ud M_{\mathrm{h}}}\,\ud M_{\mathrm{h}}\,,\label{eq:phy_sfrd}
\end{equation}
where $M_{\mathrm{h}}$ is the halo mass and $(\ud n/\ud M_{\mathrm{h}})\cdot \ud M_{\mathrm{h}}$ is the number density of haloes with masses in the range $M_{\mathrm{h}}$ to $M_{\mathrm{h}}+\ud M_{\mathrm{h}}$. In general, the SFE is a function of both redshift and halo mass \citep[see, e.g.,][]{Fialkov_2013, Mirocha_2016, Park_2019, Mittal_jwst, Dhandha_2025}. In this work we assume only a constant SFE $(f_{\star}=0.1)$ and leave the $z$- or $M_{\mathrm{h}}$-dependent models for a future work. 

Although, \citet[][hereafter press74]{PS74} is known to under-predict $\ud n/\ud M_{\mathrm{h}}$ for high halo masses \citep{Sheth_99}, computations with press74 HMF are faster because it allows to write down a closed form of collapse fraction \citep{BL04}. However, as we will see later, the difference in the resulting 21-cm signal is negligible when a more accurate form of HMF, such as \citet{Tinker_2008} is used. Hence, our fiducial choice for the dark matter HMF, $\ud n/\ud M_{\mathrm{h}}$, is the press74 form. We obtain HMF using the Python package \texttt{COLOSSUS} \citep{Diemer_2018}.

\begin{figure}
\centering
\includegraphics[width=1\linewidth]{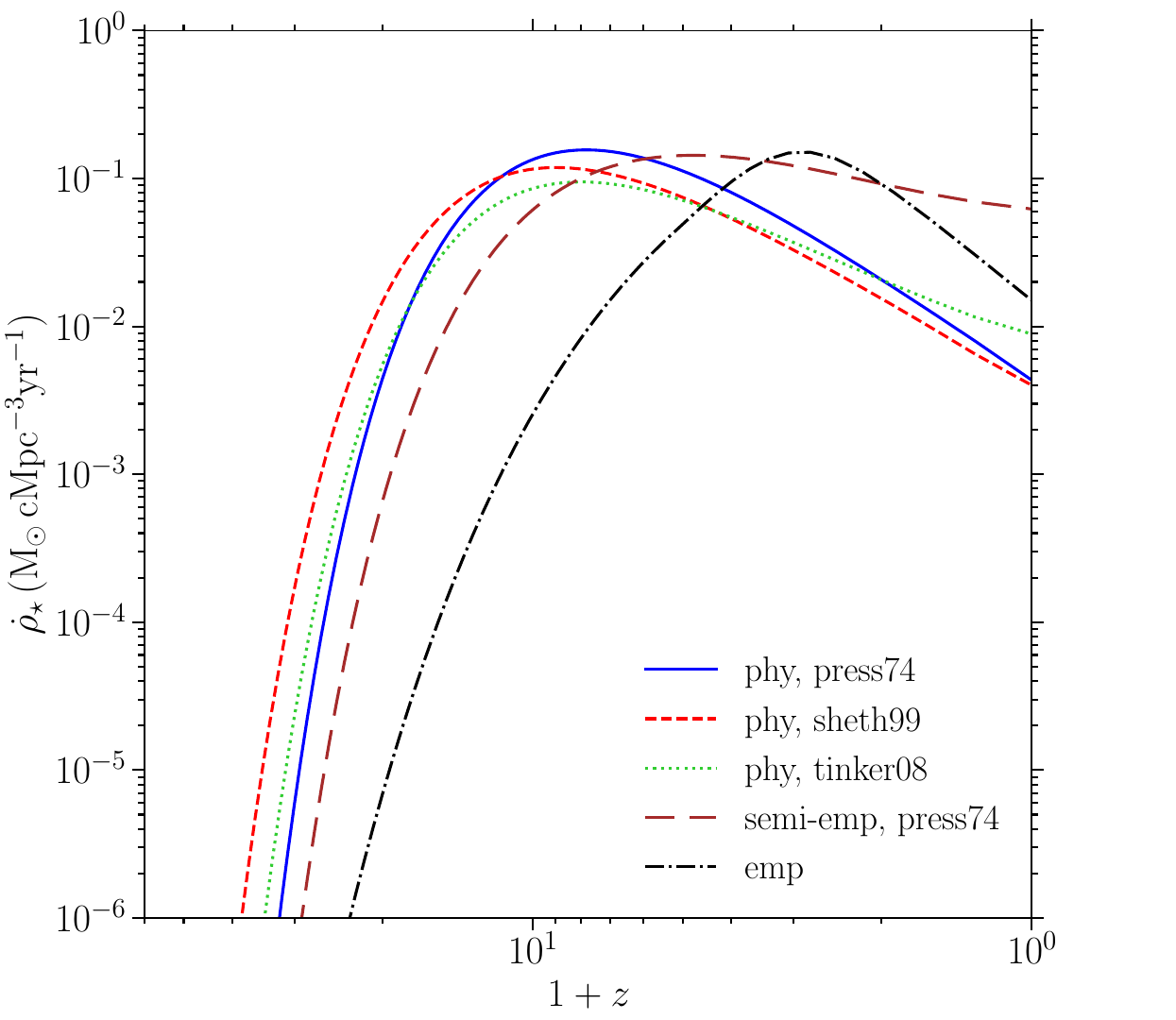}
\caption{Comoving cosmic star formation rate density in the redshift range $0\leqslant z\leqslant59$. For all physically-motivated models (equation~\ref{eq:phy_sfrd}) we adopt 10\,\% SFE and $\SI{e4}{\kelvin}$ minimum virial temperature. The solid-blue, dashed-red, and dotted-green curves correspond to Press-Schechter (`press74'), Sheth-Tormen (`sheth99'), and Tinker (`tinker08') HMF, respectively. The brown long-dashed curve is our semi-empirical model of SFRD (equation~\ref{eq:semi_sfrd}) for a press74 HMF. The dash-dotted-black curve is the empirically-motivated model of SFRD (equation~\ref{eq:emp_sfrd}).}\label{fig:sfrd}
\end{figure}

The minimum halo mass $M_{\mathrm{h,min}}$ capable of hosting stars is often parametrized by the minimum virial temperature $\mathrm{min}(T_{\mathrm{vir}})$ as \citep{BL04, DAYAL20181}
\begin{equation}
M_{\mathrm{h,min}}=10^8\frac{1}{\sqrt{\Omega_{\mathrm{m}}h^2}}\mathrm{M}_{\odot}\left[\frac{10}{1+z}\frac{0.6}{\mu}\frac{\mathrm{min}(T_{\mathrm{vir}})}{\num{1.98e4}}\right]^{3/2}\,.\label{tvir}
\end{equation}
where $\mu\approx1.22$. For the physically-motivated model $\mathrm{min}(T_{\mathrm{vir}})$ is a free parameter, fiducial choice for which is $\SI{e4}{\kelvin}$ corresponding to an atomic-cooling threshold \citep{Oh_2002}. Fig.~\ref{fig:sfrd} shows SFRD for various HMFs, namely press74 (solid-blue), \citet[][`sheth99'; dashed-red]{Sheth_99} and \citet[][`tinker08'; dotted-green]{Tinker_2008} HMF.

\begin{table*}
\centering
\caption{Free astrophysical parameters in \texttt{ECHO21}. The first column gives the parameters, second column gives a brief description of the parameter, and the last column gives the fiducial value. In addition to these parameters, the fiducial choice for SFRD model is the physically-motivated approach (equation~\ref{eq:phy_sfrd}) and the corresponding form of HMF is Press-Schechter. Due to its degeneracy with $f_{\mathrm{Ly}}$, $f_\mathrm{X}$, and $f_{\mathrm{esc}}$ we fix $f_\star$ to 0.1.}\label{tab:par}
\begin{threeparttable}
\begin{tabular}{lll}
\hline
Parameter  & Description & Fiducial value\\ \hline
\rule{0pt}{2.5ex}$f_{\mathrm{Ly}}$ & Normalization of the Ly$\upalpha$ background & $1$\\
$s$ & Spectral index of Lyman series SED\tnote{a} & $2.64$\\
$f_\mathrm{X}$ & Normalization of the X-ray background & $1$\\
$w$ & Spectral index of X-ray SED\tnote{a} & $1.5$\\
$f_{\mathrm{esc}}$ & Escape fraction of ionizing photons & 0.01\\
\makecell{min($T_{\mathrm{vir}}$)} & \makecell{Minimum virial temperature of star-forming haloes\\(For physically-motivated and semi-empirical SFRD)} & $\SI{e4}{\kelvin}$\\
\rule{0pt}{2.5ex}\makecell{$t_{\star}$} & \makecell{Characteristic star formation time-scale in\\ units of Hubble time (only for semi-empirical SFRD)} & $0.5$\\
\rule{0pt}{2.5ex}\makecell{$a$} & \makecell{$-\ud(\log_{10}\dot{\rho}_{\star})/\ud z|_{z>4}$\\(only for empirically-motivated SFRD)} & $0.257$\\ 
\hline
\end{tabular}
\begin{tablenotes}
\item[a] When the SED is expressed as energy, and not number, emitted in a unit energy range, i.e., SED$\,\propto E^{-\beta}$.
\end{tablenotes}
\end{threeparttable}
\end{table*}

\subsubsection*{Semi-empirical model}
We consider a slight variant of the physically-motivated model, which we label as the `semi-empirical' model. Following \citet{Park_2019}, an average star formation rate can be expressed as the total stellar mass $M_{\star}$ in a halo of mass $M_{\mathrm{h}}$ formed in a characteristic time-scale of $t_{\star}H^{-1}$, so that
\begin{equation}
\dot{M}_{\star} = \frac{M_{\star}}{t_{\star}H^{-1}}\,,
\end{equation}
where, as before, the stellar mass is the star formation efficiency times the gas collapsed in the haloes. Gas collapsed in the haloes in turn is just the global baryon-to-dark-matter fraction times the halo mass, so that $M_{\star}=f_{\star}(\Omega_{\mathrm{b}}/\Omega_{\mathrm{m}})M_{\mathrm{h}}$. \citet{Park_2019} additionally consider quenching of star formation by an exponential suppression factor, called the duty cycle. However, instead of assuming an exponential duty cycle, we adopt a sharp cut-off at $M_{\mathrm{h,min}}$ (defined in equation~\ref{tvir}), i.e., no star formation is possible below $M_{\mathrm{h,min}}$ \citep{Fialkov_2013}. Having obtained $\dot{M}_{\star}$ and a choice of HMF, we can write the semi-empirical SFRD (comoving) as 
\begin{equation}
\dot{\rho}_\star(z)=\frac{\Omega_\mathrm{b}}{\Omega_{\mathrm{m}}}\frac{1}{t_\star H^{-1}}\int_{M_{\mathrm{h,min}}}^{\infty} f_{\star}M_{\mathrm{h}}\frac{\ud n}{\ud M_{\mathrm{h}}}\,\ud M_{\mathrm{h}}\,,\label{eq:semi_sfrd}
\end{equation}
With the semi-empirical model we have $t_{\star}$ (default value $0.5$) as a free parameter in addition to min$(T_{\mathrm{vir}})$. As with the physically-motivated case, we consider constant SFE of $f_{\star}=0.1$. Long-dashed brown curve in Fig.~\ref{fig:sfrd} shows a semi-empirical model of SFRD for a press74 HMF, min$(T_{\mathrm{vir}})=\SI{e4}{\kelvin}$ and $t_{\star}=0.5$.

The physically-motivated and semi-empirical models have a similar steepness for high redshifts ($z\gtrsim10$) but still having predictions differing by orders of magnitude. For example long-dashed brown curve is lower than the blue curve by a factor of 10 at $z\sim20$. However, beyond their peaks, physically-motivated models drop more rapidly than semi-empirical model. By $z=0$ long-dashed brown curve is already higher than the blue curve by a factor of $\sim10$. As we will see later, different SFRD models can lead to substantial differences in the cosmic dawn 21-cm signal. 
 
\subsubsection*{Empirically-motivated model}
Besides the physically-motivated and the semi-empirical SFRD models described above we also consider an empirical SFRD model. Although, an empirical model has no theoretical backing, it is directly constrained by UV luminosity function data. Further, \texttt{ECHO21} computations with empirical model of SFRD are faster than even the physically-motivated press74 model by about an order of magnitude.

Based on the luminosity function measurements by Hubble Space Telescope (\textit{HST}) surveys, \citet{Madau_2014} proposed an empirical model for the cosmic star formation rate density.\footnote{Note that the SFRD values inferred (and hence the empirical fit to them) based on the observed luminosity function, are subject to the choice of limiting luminosity or equivalently the magnitude. As shown by \citet{Park_2019}, choosing a lower limiting luminosity leads to a higher SFRD. In this work we continue with the SFRD fit inferred based on the limiting magnitude adopted by many recent studies \citep[e.g.][]{Donnan_2023, Donnan_2024}.} Their model explains the low-redshift SFRD quite well. However, it overestimates the SFRD for $z\gtrsim8$; recent \textit{JWST} observations reveal that SFRD has an exponential drop rather than a power-law drop in the range $8\lesssim z\lesssim13$ \citep{Donnan_2023, Harikane_2023, Donnan_2024}. Observations by \textit{HST} in the pre-\textit{JWST} era have also hinted towards this behaviour \citep[e.g.,][]{Mcleod_2016}. Above $z\sim13$, the SFRD trend is much more uncertain because of the lack of spectroscopic confirmation of galaxy candidates observed by \textit{JWST}. Although, there is now a growing evidence towards an even sharper decline in SFRD than an exponential drop above $z\sim13$ \citep[e.g.][]{Robertson_2024}, further corroboration is needed.

Given the above discussion, we assume a Madau-Dickinson form at low redshifts while an exponential form at high redshifts. Thus, our empirical SFRD (comoving) is given as
\begin{equation}
\dot{\rho}_\star(z)=
\begin{dcases}
\frac{0.015(1+z)^{2.73}}{1+\left[(1+z)/3\right]^{6.24}} & \text{if } 0\leqslant z\leqslant 4\,,\\
\dot{\rho}_\star(z=4)\cdot 10^{-a(z-4)} & \text{if } z>4\,.
\end{dcases}
\label{eq:emp_sfrd}
\end{equation}
in units of $\mathrm{M_{\odot}yr^{-1}cMpc^{-3}}$. We consider one free parameter here: $a$, which we define as
\begin{equation}
a = -\left.\frac{\ud\log_{10}\dot{\rho}_{\star}}{\ud z}\right|_{z>4}
\end{equation}
fiducial value for which is $0.257$ \citep{Gupta_2025}. Note that we do not consider any free parameter associated with the low-redshift SFRD as it accurately explains observations in this redshift range. The dash-dotted black curve in Fig.~\ref{fig:sfrd} shows the empirically-motivated SFRD with the fiducial value of $a$.

Just as in the case of a physically-motivated model, we assume the star formation with the rate density given by equation~\eqref{eq:emp_sfrd} to begin at $z=59$. Thus, we assume the exponential form continues to be valid at high redshifts.\\
[1\baselineskip]

This completes our discussion of IGM modelling and the required ingredients. Table~\ref{tab:par} summarises our free astrophysical parameters and their fiducial values. Note that we do not consider star formation efficiency ($f_{\star}$) to be a free parameter since it is degenerate with $f_{\mathrm{Ly}}, f_{\mathrm{X}}$, and $f_{\mathrm{esc}}$. Similarly, we have fixed the yield of ionizing photons ($I_{\mathrm{ion}}$) as it is degenerate with $f_{\mathrm{esc}}$.

\section{Results}\label{sec:res}
In this section we will look at the results generated by our code \texttt{ECHO21}. In section~\ref{sec:fid} we look at a single realization of the global signal for our fiducial choice of parameters. Here we also show the evolution of electron fraction, volume-filling factor, and gas temperature from dark ages until the end of epoch of reionization. In section~\ref{sec:sfrd_impact} we investigate the impact of the choice of SFRD model on the global signal. In section~\ref{sec:para} we explore a large parameter space of astrophysical and cosmological parameters.

\subsection{Fiducial model}\label{sec:fid}
We first show results for our fiducial model for which $f_{\mathrm{Ly}}=1, s=2.64, f_{\mathrm{X}}=1, w=1.5, f_{\mathrm{esc}}=0.01$, and min($T_{\mathrm{vir}})=\SI{e4}{\kelvin}$. In addition to these we choose the physically-motivated model of SFRD with the Press-Schechter form of HMF (see Table~\ref{tab:par} for a summary).

\begin{figure}
\centering
\includegraphics[width=1\linewidth]{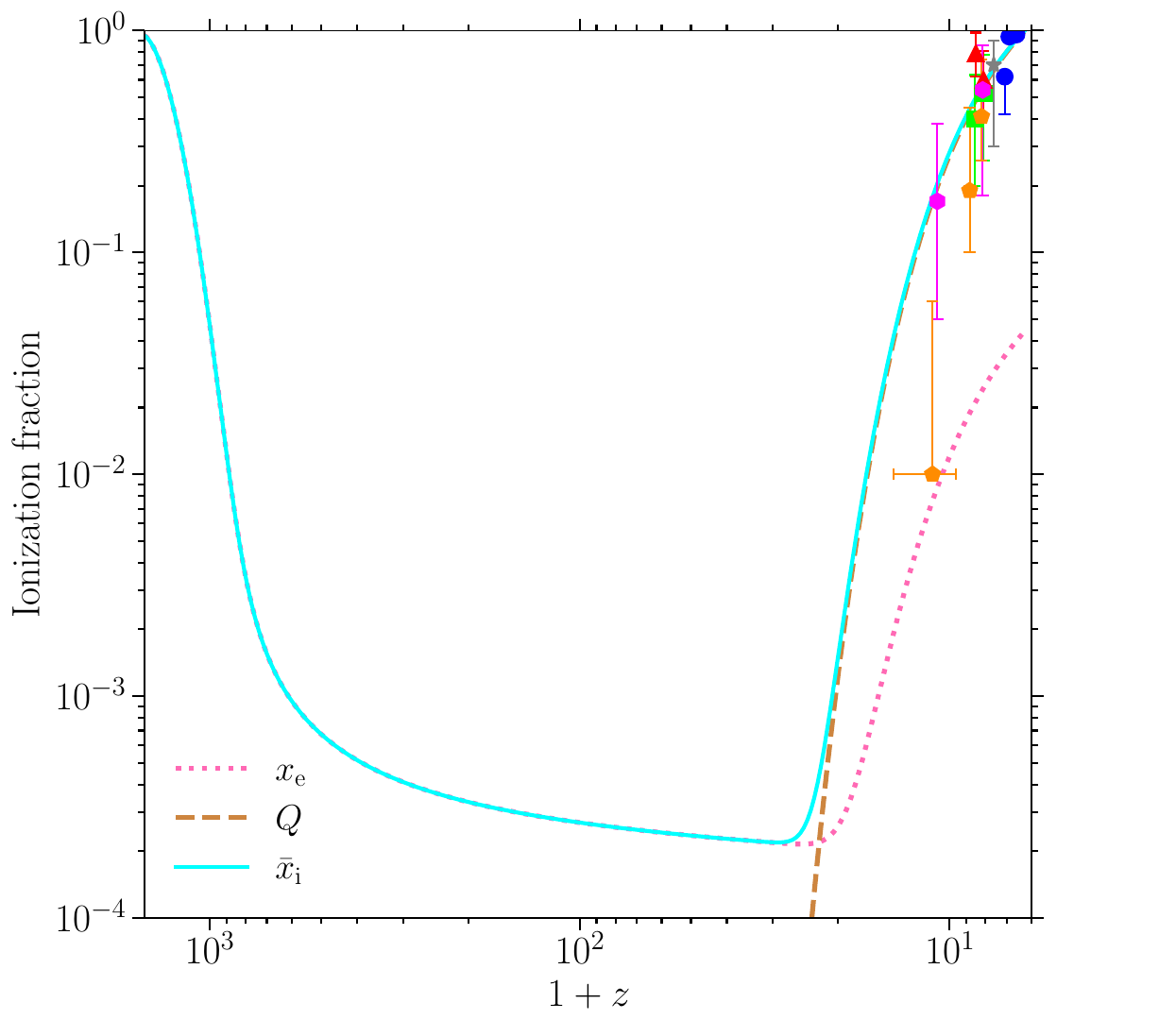}
\caption{Bulk IGM electron fraction ($x_{\mathrm{e}}$, dotted pink), volume-filling factor ($Q$, dashed brown), and two-zone-model-averaged ionization fraction ($\bar{x}_{\mathrm{i}}$, solid cyan) computed using \texttt{ECHO21}. In this fiducial model, 50\% reionization is complete by $7.42$ and 100\% by $z=5.3$. We also show constraints on the state of ionization from some of the recent literature, namely \citet[][blue circles]{McGreer_2014}, \citet[][red triangles]{Greig_2016, Greig_2019}, \citet[][green squares]{Davies_2018}, \citet[][grey star]{Ning_2022}, \citet[][orange pentagons]{Nakane_2024}, and \citet[][magenta hexagons]{Umeda_2024}.}\label{fig:ion}
\end{figure}

Fig.~\ref{fig:ion} shows the trend for the bulk IGM electron fraction ($x_{\mathrm{e}}$), volume-filling factor ($Q$) and the two-zone-model-averaged ionization fraction ($\bar{x}_{\mathrm{i}}$) with redshift. The trend for $x_{\mathrm{e}}$ in the dark ages agrees with the standard recombination calculations \citep[see, e.g.,][]{Peebles_1968, Seager_1999, Seager_2000}. In the absence of external ionizing sources $x_{\mathrm{e}}$ would saturate to $\sim2\times10^{-4}$ by $z\sim5$. However, X-rays emitted in the comic dawn period with their long mean free path can penetrate deep into the IGM and cause ionizations resulting in a rise in $x_{\mathrm{e}}$. For our fiducial model we see this for $z\lesssim22$.

Fig.~\ref{fig:ion} also shows the progression of epoch of reionization by the dashed-brown curve. The redshift of reionization (when the Universe was 50\% ionized) is $z_{\mathrm{re}}=7.42$ (which is within $1\sigma$ uncertainty bars of \textit{Planck} 2018 results; $7.64\pm0.74$) and the redshift when the Universe is fully ionized is $z=5.30$, in agreement with large Ly$\upalpha$ opacity fluctuations and low CMB optical depth data \citep{Kulkarni}. In the same figure we also show constraints on the reionization from various probes and techniques. These include dark gaps in the Ly$\upalpha$ forest \citep{McGreer_2014}, quasar damping wings \citep{Greig_2016, Davies_2018, Greig_2019}, the effective Ly$\upalpha$ opacity of the IGM \citep{Ning_2022}, Ly$\upalpha$ emission equivalent widths \citep{Nakane_2024}, and galaxy damping wings \citep{Umeda_2024}. As evident, our fiducial model of reionization (represented by dashed-brown curve) is in reasonable agreement with inferences from various observations.

Finally, relevant to the calculation of 21-cm signal (equation~\ref{eq:DeltaT}) we also show the two-zone-model-averaged ionization fraction, $\bar{x}_{\mathrm{i}}$, defined in equation~\eqref{eq:gaif}. See the solid cyan-coloured curve.

\begin{figure}
\centering
\includegraphics[width=1\linewidth]{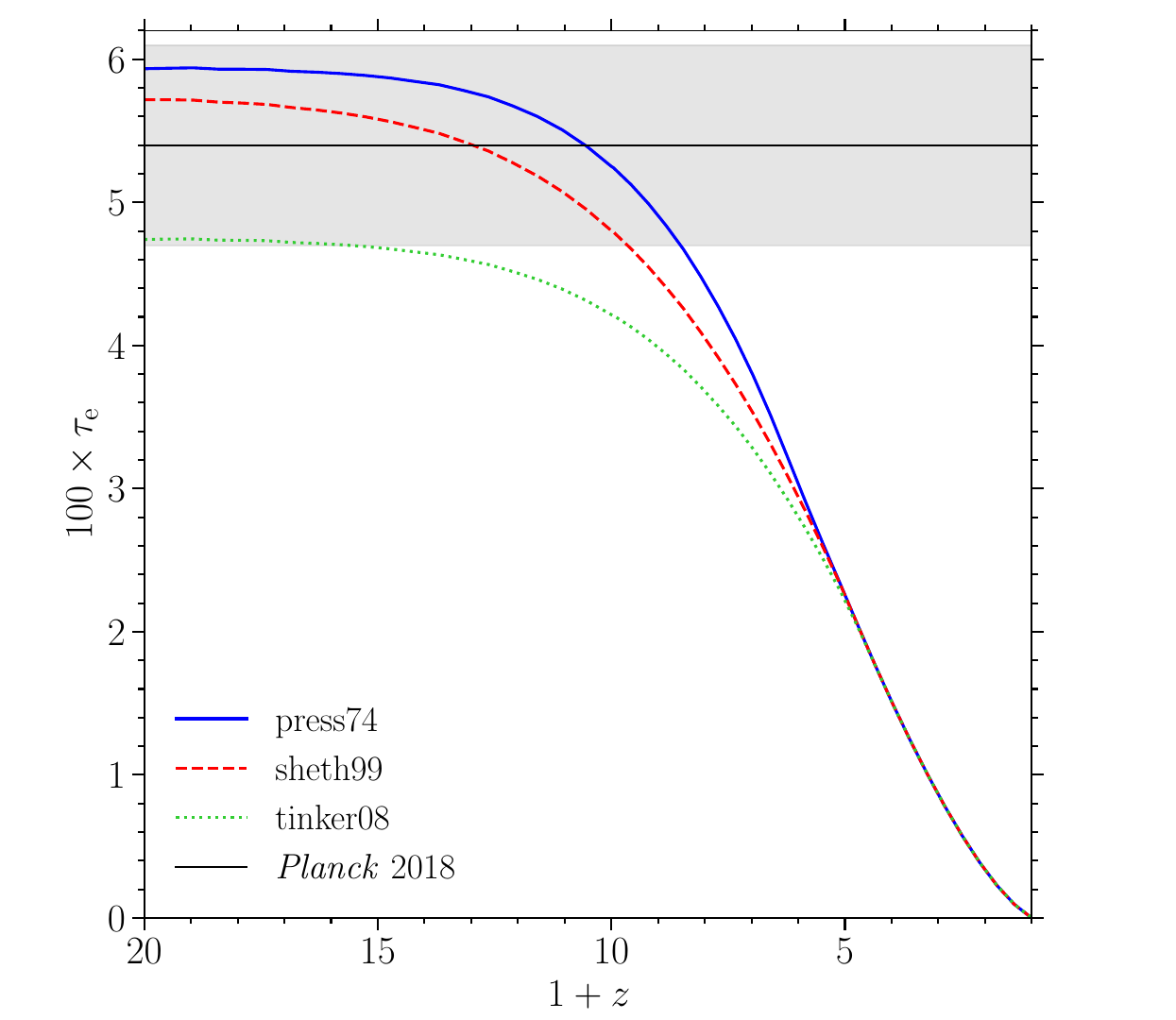}
\caption{Electron-scattering optical depth as a function of maximum integration redshift $z$ for our fiducial model generated using \texttt{ECHO21}. For all three cases we work with a physically-motivated SFRD model. The optical depth for press74 HMF out to the beginning of cosmic dawn comes out to be $0.0592$ which is within 1$\sigma$ limits of \textit{Planck} 2018 results.}\label{fig:tau}
\end{figure}

Another probe of EoR is the electron-scattering optical depth. In Fig.~\ref{fig:tau} we show the electron-scattering optical depth as a function of maximum integration redshift $z$. For our fiducial model and the default HMF choice (press74) the total optical depth (by setting $z=z_{\star}$ in equation~\ref{eq:tau}) comes out to be $\tau_{\ue}\approx0.0592$ which is within $1\sigma$ uncertainty bars of \textit{Planck} 2018 results; $0.054\pm0.007$. See the solid blue-coloured curve. We also show $\tau_\ue$ for other HMFs, namely sheth99 (dashed red) and tinker08 (dotted green) still assuming a physically-motivated model of SFRD. We note that the total optical depth for other HMFs is within $1\sigma$ uncertainty bars of \textit{Planck} 2018 results as well.

\begin{figure*}
\centering
\includegraphics[width=1\linewidth]{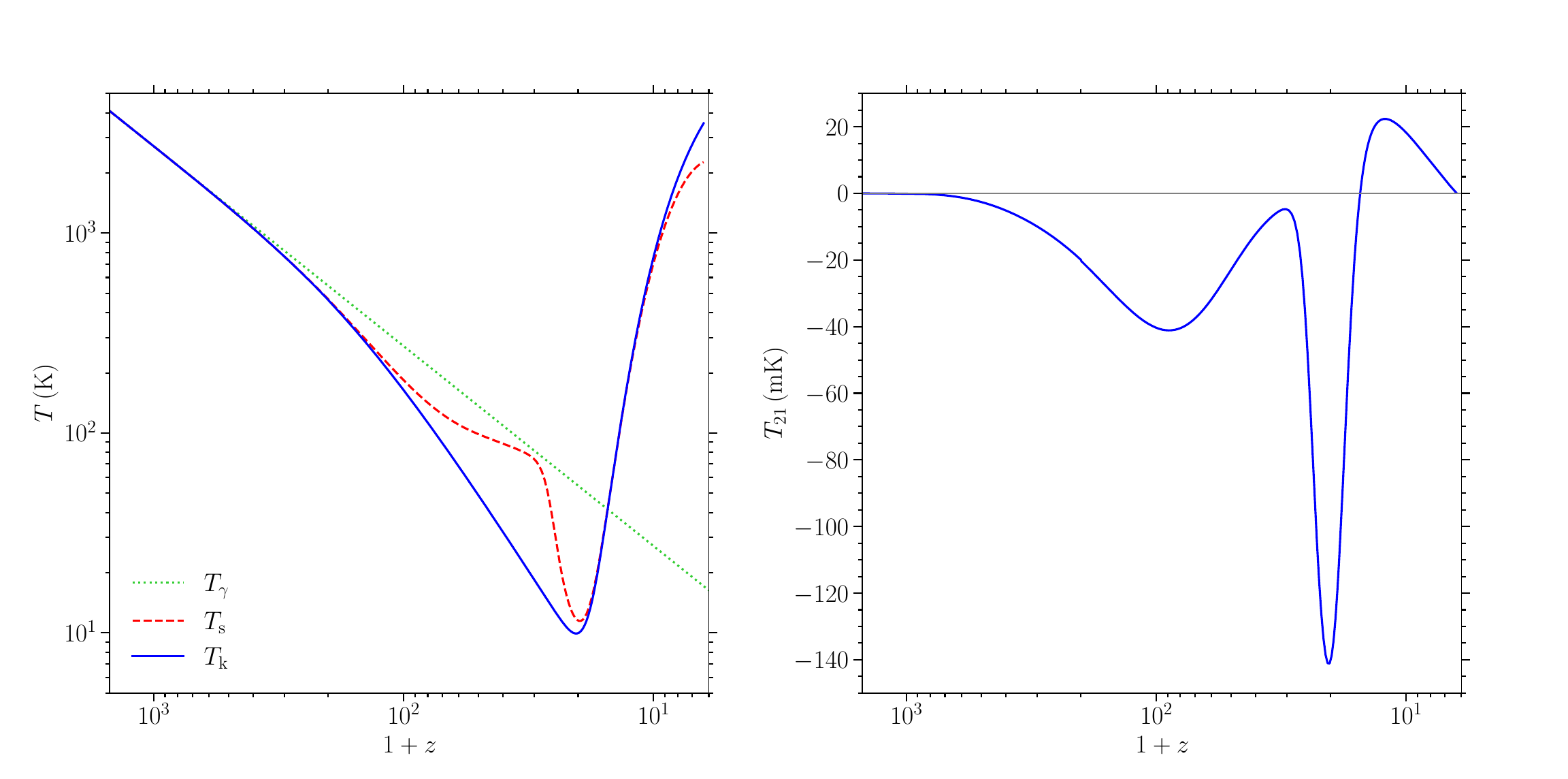}
\caption{Left: gas kinetic temperature (solid blue), CMB temperature (dotted green), and spin temperature (dashed red). Right: the global 21-cm signal. For fiducial set of parameters (see Table~\ref{tab:par}) the strongest signal is $\SI{-142.4}{\milli\kelvin}$, observed at $z = 19.21$. The redshift range is $5\leqslant z\leqslant1500$, covering dark ages until the end of EoR.}\label{fig:t21}
\end{figure*}

Fig.~\ref{fig:t21} shows the main results of our work. Left panel shows the gas, CMB and the spin temperature as a function of redshift and the right panel shows the corresponding global 21-cm signal, which is the main product of \texttt{ECHO21}. Although, we set our starting of cosmic dawn at $z=59$, gas temperature is seen to rise (due to X-ray and Ly$\upalpha$ heating) only after $z\sim19.5$ before which it drops adiabatically. Still, the effect of star formation can be seen as early as $z=29.3$ when the Ly$\upalpha$ photons by the newly formed stars start to render the signal in absorption. At this stage the global signal is \SI{-4.7}{\milli\kelvin}.

The first absorption peak (in the dark ages) is $\SI{-41.1}{\milli\kelvin}$ observed at $z=87.5$. This absorption is caused by hydrogen-electron and hydrogen-hydrogen collisions. The second absorption peak (in the cosmic dawn) is $\SI{-142.4}{\milli\kelvin}$ observed at $z=19.2$. This absorption is dominantly induced by the Ly$\upalpha$ photons. The maximum emission signal (in the epoch of reionization) is $\SI{22.2}{\milli\kelvin}$ at $z=11.2$. Once the Universe becomes completely ionized by $z=5.3$, the signal disappears.

\begin{figure}
\centering
\includegraphics[width=1\linewidth]{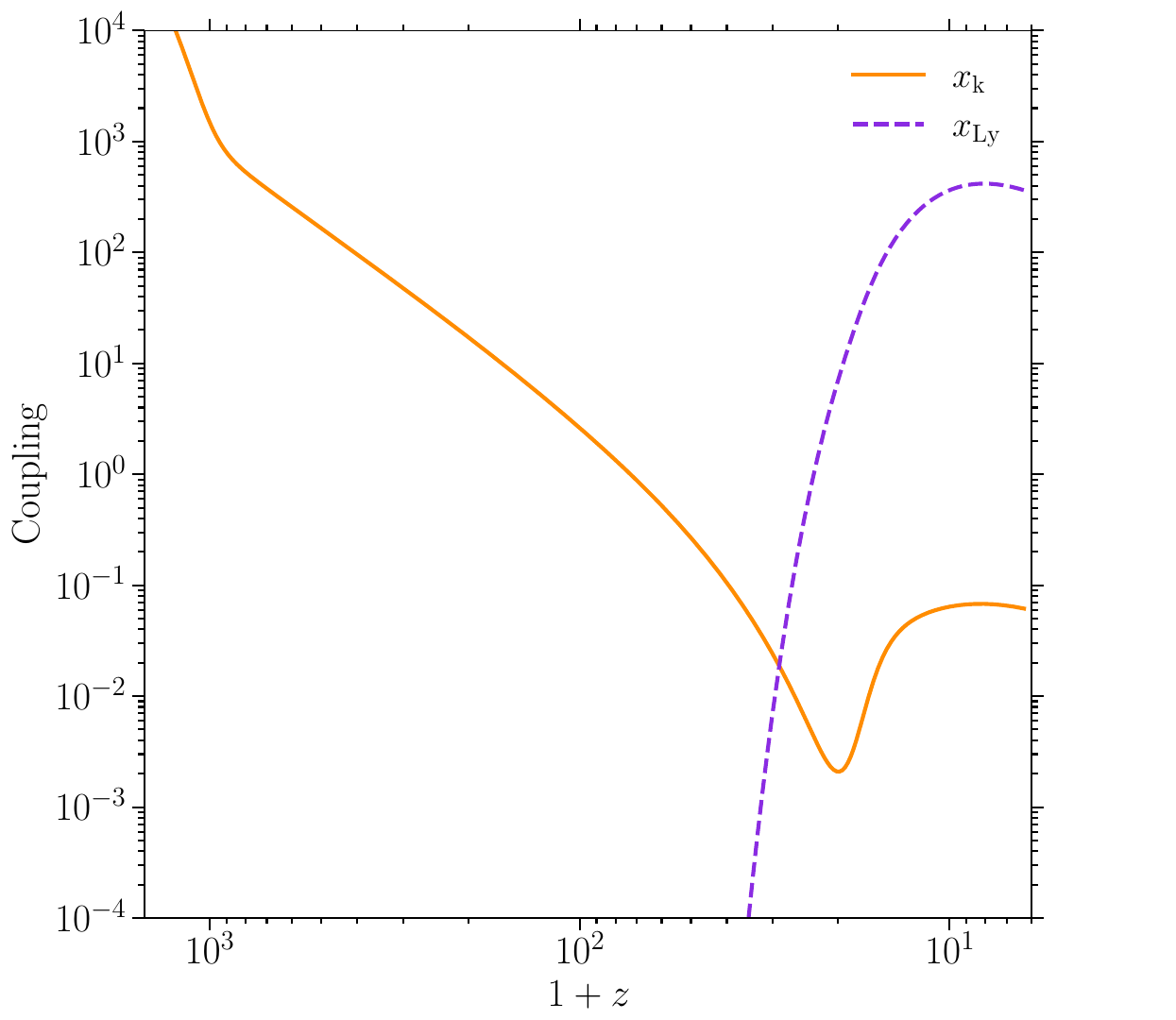}
\caption{Collisional (solid orange) and Ly$\upalpha$ (dashed purple) coupling for our fiducial set of parameters. Fig.~\ref{fig:t21} shows the corresponding spin temperature and 21-cm signal.}\label{fig:coup}
\end{figure}

Fig.~\ref{fig:coup} shows the trends of couplings with redshift. The collisional coupling is high in the dark ages but drops rapidly as the gas temperature and gas density drop with the expanding Universe. However, we see a small rise in $x_{\mathrm{k}}$ at $z\approx19$ as a result of rising temperature due to gas heating. But eventually the effect of rapid Hubble expansion at low redshifts, and hence density dilution, dominates over temperature rise and again $x_{\mathrm{k}}$ continues to drop. The dashed-red curve shows the Ly$\upalpha$ coupling which is only relevant in the cosmic dawn period ($z<z_\star$). Since $x_{\mathrm{Ly}}$ tracks the star formation, $x_{\mathrm{Ly}}$ rises rapidly for $z\gtrsim6.5$ whereafter it starts to drop because of the drop in star formation rate at $z\approx6.5$. However, throughout $z<27$, $x_{\mathrm{Ly}}$ dominates over $x_{\mathrm{k}}$ by several orders of magnitude.

\begin{figure}
\centering
\includegraphics[width=1\linewidth]{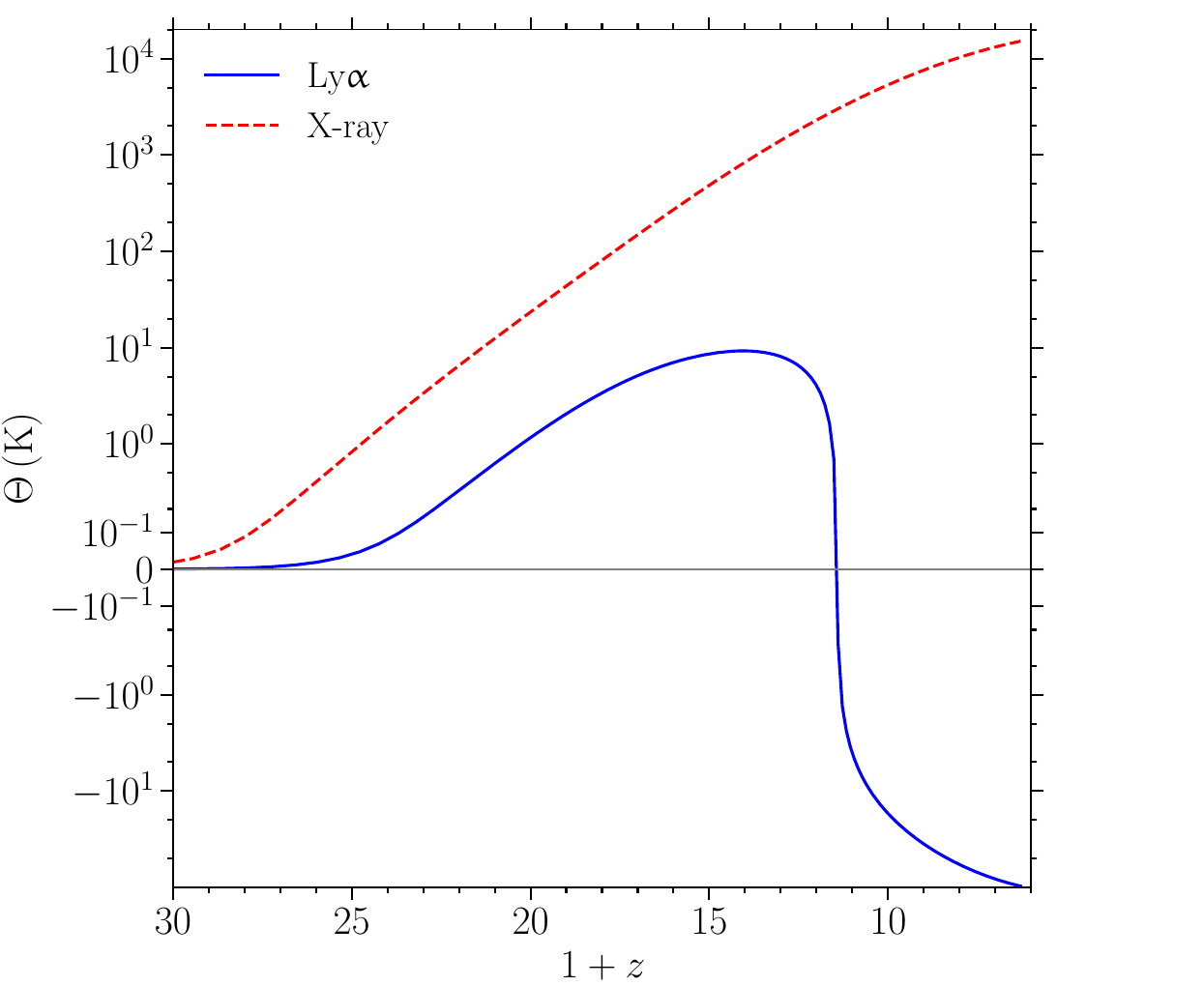}
\caption{Comparison of Ly$\upalpha$ (solid blue) and X-ray (dashed red) heating in the redshift range $5\leqslant z\leqslant29$ for fiducial set of parameters. $y$ axis shows the quantity $\Theta=2q/(3n_{\mathrm{b}}k_{\mathrm{B}}H)$, which has the units of temperature. Although X-ray heating is dominant over Ly$\upalpha$ heating throughout the range; Ly$\upalpha$ heating is an important term to consider for an accurate modelling of gas temperature evolution.}\label{fig:heat}
\end{figure}

Fig.~\ref{fig:heat} shows the trends of Ly$\upalpha$ and X-ray heating for our fiducial model. For our fiducial choice of parameters, X-ray heating dominates over Ly$\upalpha$ heating for the majority of redshift range. However, for wide space of parameters, high $f_{\mathrm{Ly}}$ and low $f_{\mathrm{X}}$, Ly$\upalpha$ heating can be comparable to X-ray heating. A notable consequence of this is that for low $f_{\mathrm{X}}$ values, the signal strength does not necessarily scale with $f_{\mathrm{Ly}}$ monotonically as shown in the past work \citep[e.g.][]{Pritchard_2010}. This is because the Ly$\upalpha$ heating and coupling have opposing impacts on the signal strength. This implies that in low X-ray heating regime and with all other parameters remaining fixed, there should be an optimum value of $f_{\mathrm{Ly}}$ to achieve the deepest signal at cosmic dawn.

\begin{figure}
\centering
\includegraphics[width=1\linewidth]{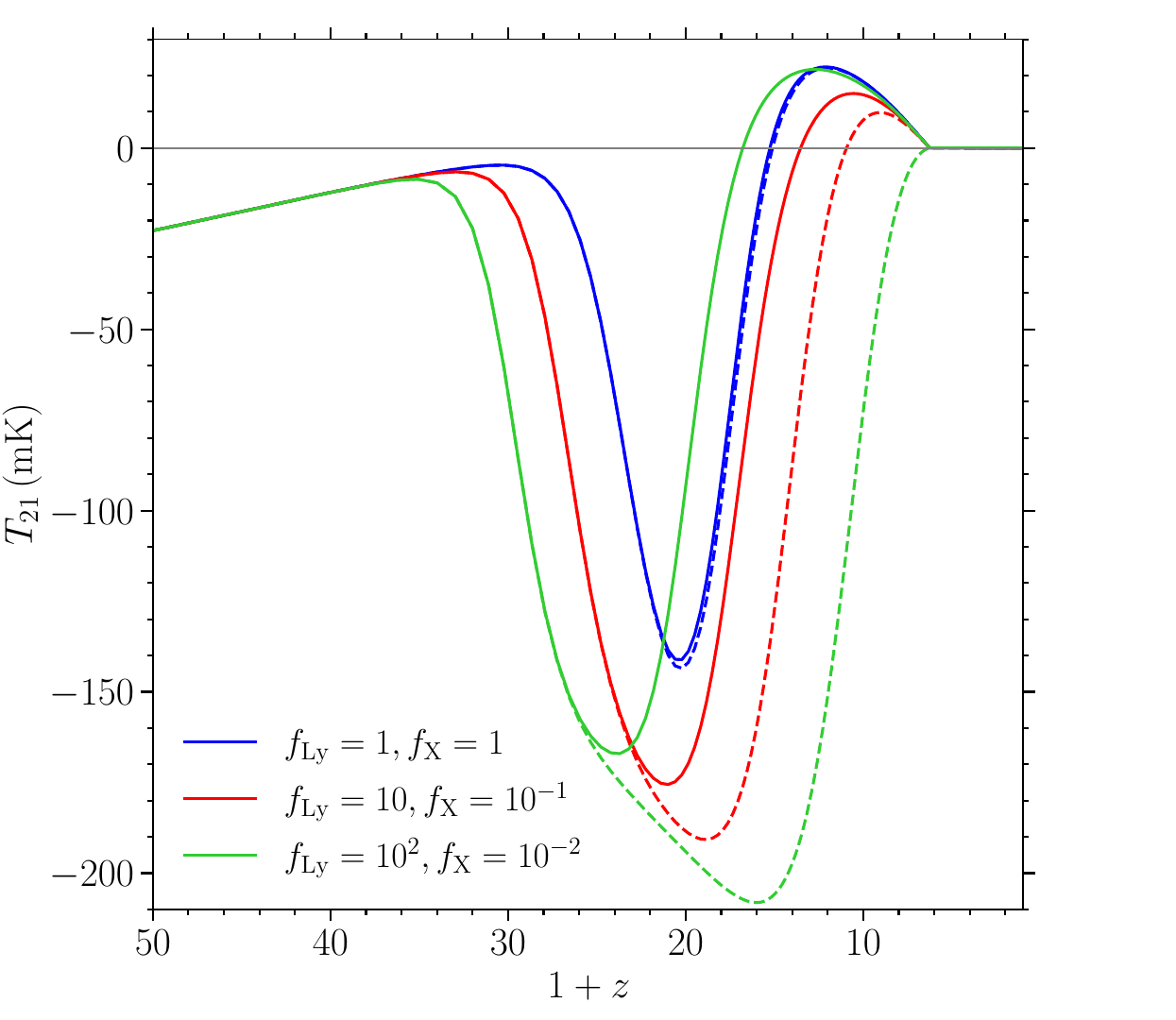}
\caption{Solid and dashed curves include and exclude the Ly$\upalpha$ heating, respectively. Everything else is the same for the same colour pair. Different colours correspond to different parameter set as labelled in the legends. Evidently, Ly$\upalpha$ heating is an important term to consider for an accurate modelling of gas temperature evolution.}\label{fig:lya}
\end{figure}

Existing codes such as \texttt{ARES} or \texttt{ZEUS21} do not account for Ly$\upalpha$ heating. For a low X-ray emissivity and strong Ly$\upalpha$ emissivity, not accounting for Ly$\upalpha$ heating can introduce significant biases. We demonstrate this in Fig.~\ref{fig:lya} for some representative cases. The dashed and solid curves for each colour follow the same modelling and parameters except for the Ly$\upalpha$ heating. Dashed curves do not account for Ly$\upalpha$ heating and hence lead to a stronger 21-cm signal compared to the solid curves. Thus, we emphasise that as part of the standard model of IGM, Ly$\upalpha$ heating is a crucial component.

Finally, we note that while X-ray photons always cause heating of IGM, Ly$\upalpha$ photons can produce cooling effect as well. For our choice of parameters this is observed for $z\lesssim10.5$. The continuum Ly$\upalpha$ photons always lead to heating but the cooling is brought about by the injected Ly$\upalpha$ photons. After $z\approx10.5$, we find that $J_{\mathrm{Ly}}^\mathrm{c}I_\mathrm{c}<|J_{\mathrm{Ly}}^\mathrm{i} I_\mathrm{i}|$ (see equation~\ref{eq:lyaheat}) and thus $q_{\mathrm{Ly}}<0$, i.e., we get a net cooling effect. 

\subsection{Impact of SFRD modelling on the 21-cm signal}\label{sec:sfrd_impact}
\begin{figure}
\centering
\includegraphics[width=1\linewidth]{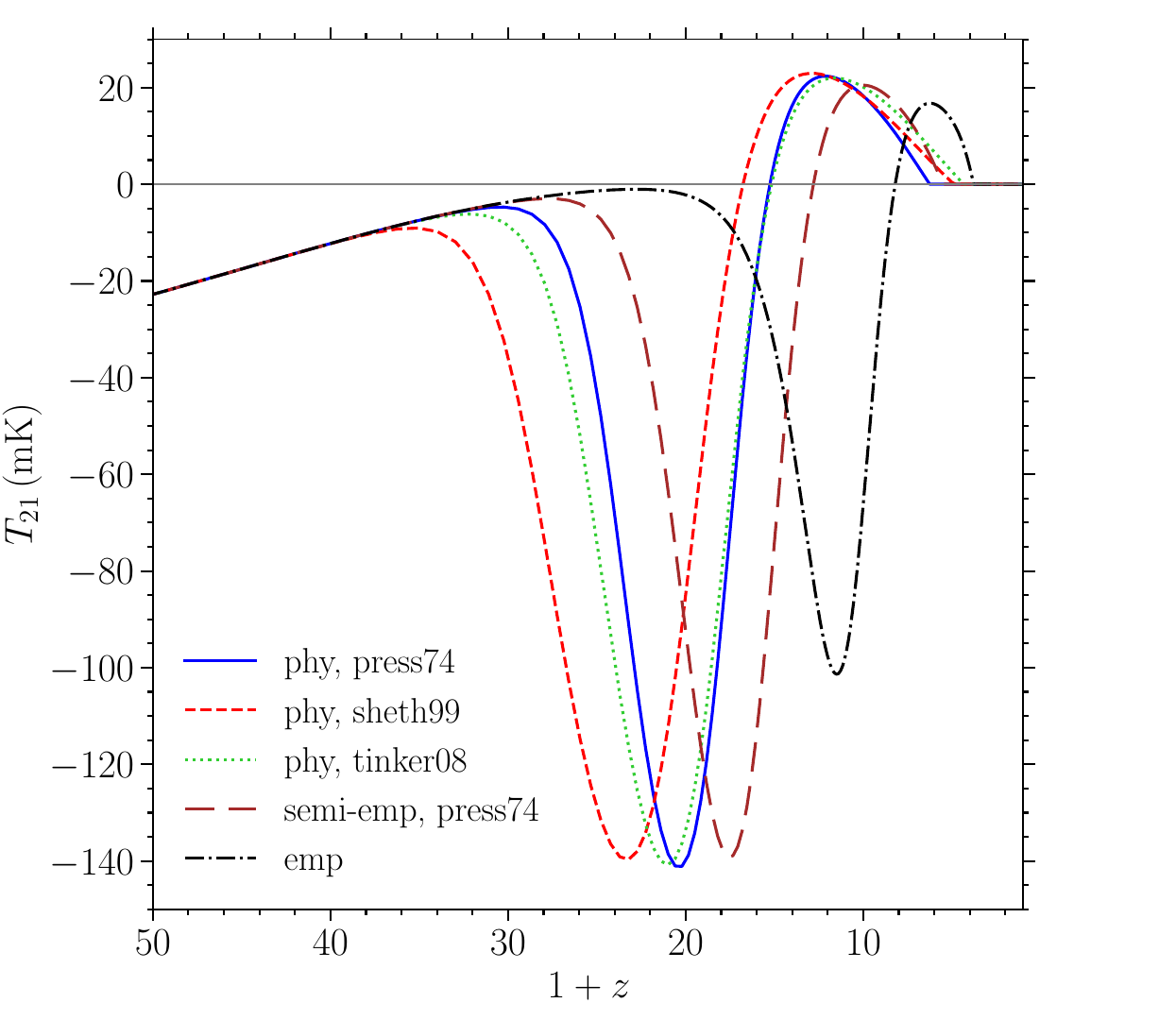}
\caption{Impact of SFRD on the 21-cm signal. Solid-blue curve (repeated from Fig.~\ref{fig:t21}) corresponds to a \citet{PS74} HMF (`press74'), dashed-red curve corresponds to a \citet{Sheth_99} HMF (`sheth99'), dotted-green curve corresponds to a \citet{Tinker_2008} HMF (`tinker08'), long-dashed-brown signal corresponds to a semi-empirical SFRD model for press74 HMF (equation~\ref{eq:semi_sfrd}) and dash-dotted-black curve corresponds to an empirical SFRD model (equation~\ref{eq:emp_sfrd}). Fig.~\ref{fig:sfrd} shows the SFRD for these models.}\label{fig:t21_sfrd}
\end{figure}

In this section we explore the impact of SFRD modelling on the 21-cm signal. As described in section~\ref{sec:res} we consider three types of SFRD models, namely a physically-motivated, semi-empirical, and an empirically-motivated model. Physically-motivated and semi-empirical models need a choice of HMF. In Fig.~\ref{fig:t21_sfrd} we show representative cases of 21-cm signal corresponding to press74 (fiducial), \citet[][`sheth99']{Sheth_99}, and \citet[][`tinker08']{Tinker_2008} HMFs. Since press74 underestimates SFRD (for $z\gtrsim11.8$, for fiducial model parameters), the activation of Ly$\upalpha$ coupling is delayed and so is the 21-cm signal compared to other cases of HMFs. Accordingly, the difference between deepest feature in the solid-blue and dashed-red curve is $\sim\SI{1.7}{\milli\kelvin}$ and a difference of $3$ on the redshift axis. From these results it is evident that the choice of HMF is important in the modelling of the 21-cm signal. Conversely, given some observational data the choice of HMF will impact the astrophysical parameters inferred. See also \citet{Greig_2024}, who investigate both these aspects in more detail. Our results on the impact of the choice of the HMF on the 21-cm signal modelling qualitatively agree with \citet{Greig_2024}.

For all physically-motivated models we use the same set of cosmological and astrophysical parameters; the fiducial set (see Table~\ref{tab:par}). Also, note that we used a constant SFE of 10\% for all the models. We plan to add a more flexible SFE model in an upcoming version of \texttt{ECHO21}.

In addition to the physically-motivated models for SFRD we also show 21-cm signal using a semi-empirical and an empirically-motivated model for SFRD (equations~\ref{eq:semi_sfrd} and \ref{eq:emp_sfrd}, respectively). Because of the late start of star formation for a semi-empirical model and more so for the empirically-motivated SFRD (see long-dashed-brown and dash-dotted-black curve in Fig.~\ref{fig:sfrd}, respectively), the signal is delayed with respect to the physically-motivated models. The deepest feature in the long-dashed-brown curve is shallower by $\sim\SI{1.9}{\milli\kelvin}$ and delayed by $2.6$ on the redshift axis compared to the solid-blue curve. The deepest feature in the dash-dotted-black curve is shallower by $\sim\SI{39.4}{\milli\kelvin}$ and delayed by $8.8$ on the redshift axis compared to solid-blue curve.

Previous works have considered the impact of SFE modelling \citep[e.g.][]{Fialkov_2013, Mirocha_2016,Park_2019}, resolving and modelling individual galaxy formation histories \citep{Mirocha_2021b}, including feedback-impacted Pop III star formation \citep[e.g.][]{Munoz_2022, Cruz_2025}, HMF modelling \citep{Greig_2024} on the 21-cm signal. However, these effects have not been studied across different SFRD models (see section~\ref{sec:sfrd}). In this work, we have demonstrated that the choice of SFRD model can lead to substantial differences in the signal predicted.

\subsection{Exploring a large parameter space}\label{sec:para}
In section~\ref{sec:fid} we showed a single instance of global signal for fiducial set of parameters. Here we generate global signals for a large number of astrophysical parameters. We choose five equispaced points for all six astrophysical parameters (for physically-motivated model of SFRD) listed in Table~\ref{tab:par} so that we have a total of 15625 models. The range is $[10^{-2},10^2], [0,3], [10^{-2},10^2], [0,3], [10^{-3},1.0]$, and $[10^2,10^6]\,$K for $f_{\mathrm{Ly}}, s, f_{\mathrm{X}}, w, f_{\mathrm{esc}}$, and min$(T_{\mathrm{vir}})$, respectively. We consider $f_{\mathrm{Ly}}, f_{\mathrm{X}}, f_{\mathrm{esc}}$ and min$(T_{\mathrm{vir}})$ to be log-spaced and $s$ and $w$ to be linearly spaced. On 50 CPUs, 15625 signals are generated in approximately 17\,minutes.

\begin{figure}
\centering
\includegraphics[width=1\linewidth]{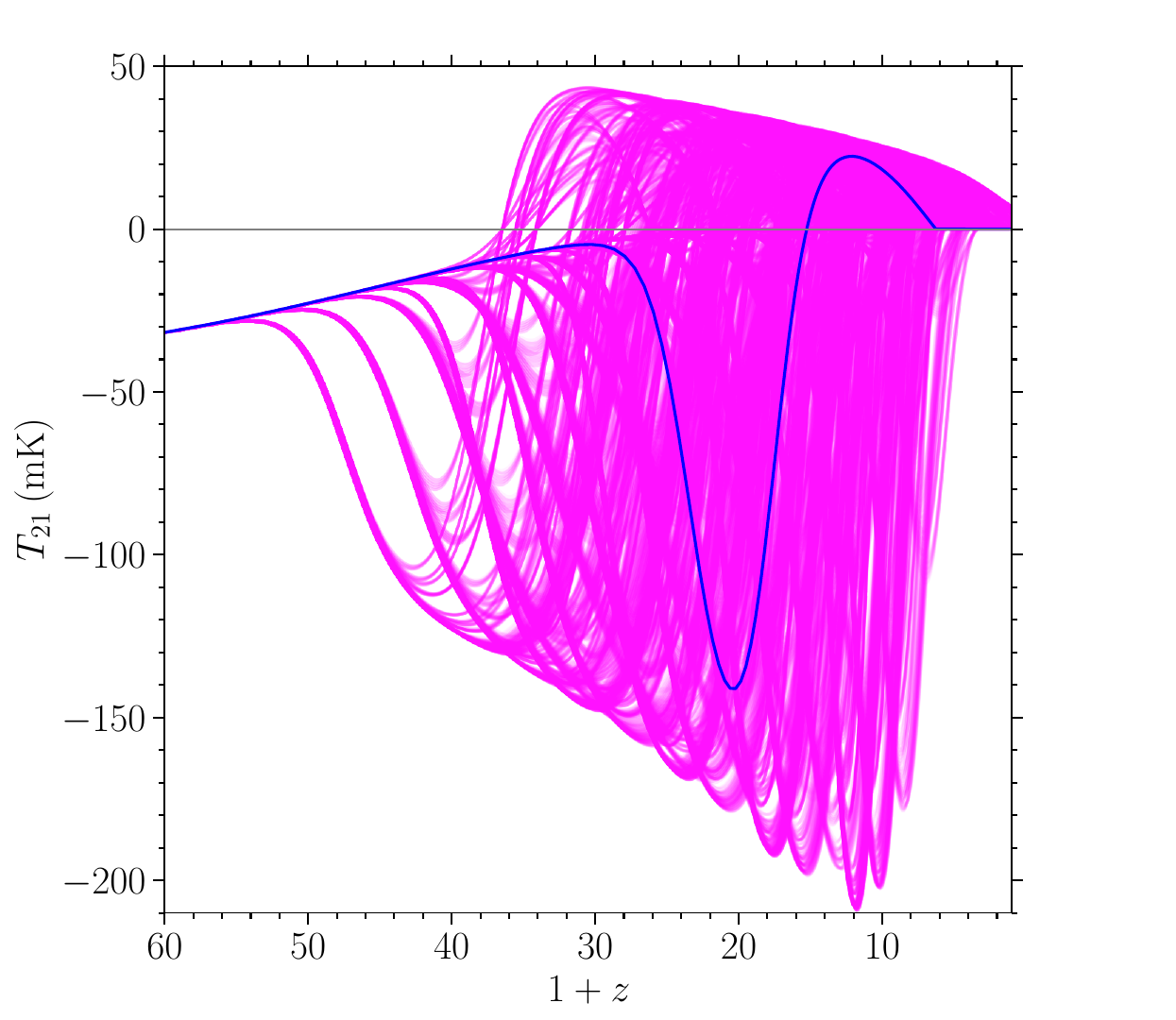}
\caption{15625 global signals (in magenta) generated with \texttt{ECHO21} for different astrophysical parameters; we vary $f_{\mathrm{Ly}}, s, f_{\mathrm{X}}, w, f_{\mathrm{esc}}$, and min$(T_{\mathrm{vir}})$. See text for the range of values we choose. The cosmological parameters are fixed at the median values quoted at the end of introductory section. The solid-blue curve shows the fiducial signal (repeated from Fig.~\ref{fig:t21}).}\label{fig:large}
\end{figure}

Fig.~\ref{fig:large} shows 15625 global signals (magenta-coloured curves) generated using \texttt{ECHO21} for a large astrophysical parameter space. The cosmological parameters are fixed at their median values mentioned in the introductory section. The solid-blue curve shows the fiducial model repeated from Fig.~\ref{fig:t21}. Note that we show the signals only in the cosmic dawn and EoR range ($0\leqslant z\leqslant59$) as the astrophysical processes do not impact the dark ages signal. Thus, in this case the dark ages signal is common for all the models.

We find that the signal responds most strongly to parameters $f_{\mathrm{Ly}}, f_{\mathrm{X}}, w$, and min$(T_{\mathrm{vir}})$, mildly to $f_{\mathrm{esc}}$ and negligibly to $s$. In general, increasing $f_{\mathrm{Ly}}$ shifts the signal to earlier epochs and deepens the signal (see also the discussion in subsection~\ref{sec:fid}), increasing $f_{\mathrm{X}}$ shifts the signal to earlier epochs and weakens the absorption signal, increasing $w$ delays and deepens the absorption signal, increasing $f_{\mathrm{esc}}$ leads to an earlier disappearance of the signal due to a rapid reionization, and increasing min$(T_{\mathrm{vir}})$ delays the signal.

\begin{figure}
\centering
\includegraphics[width=1\linewidth]{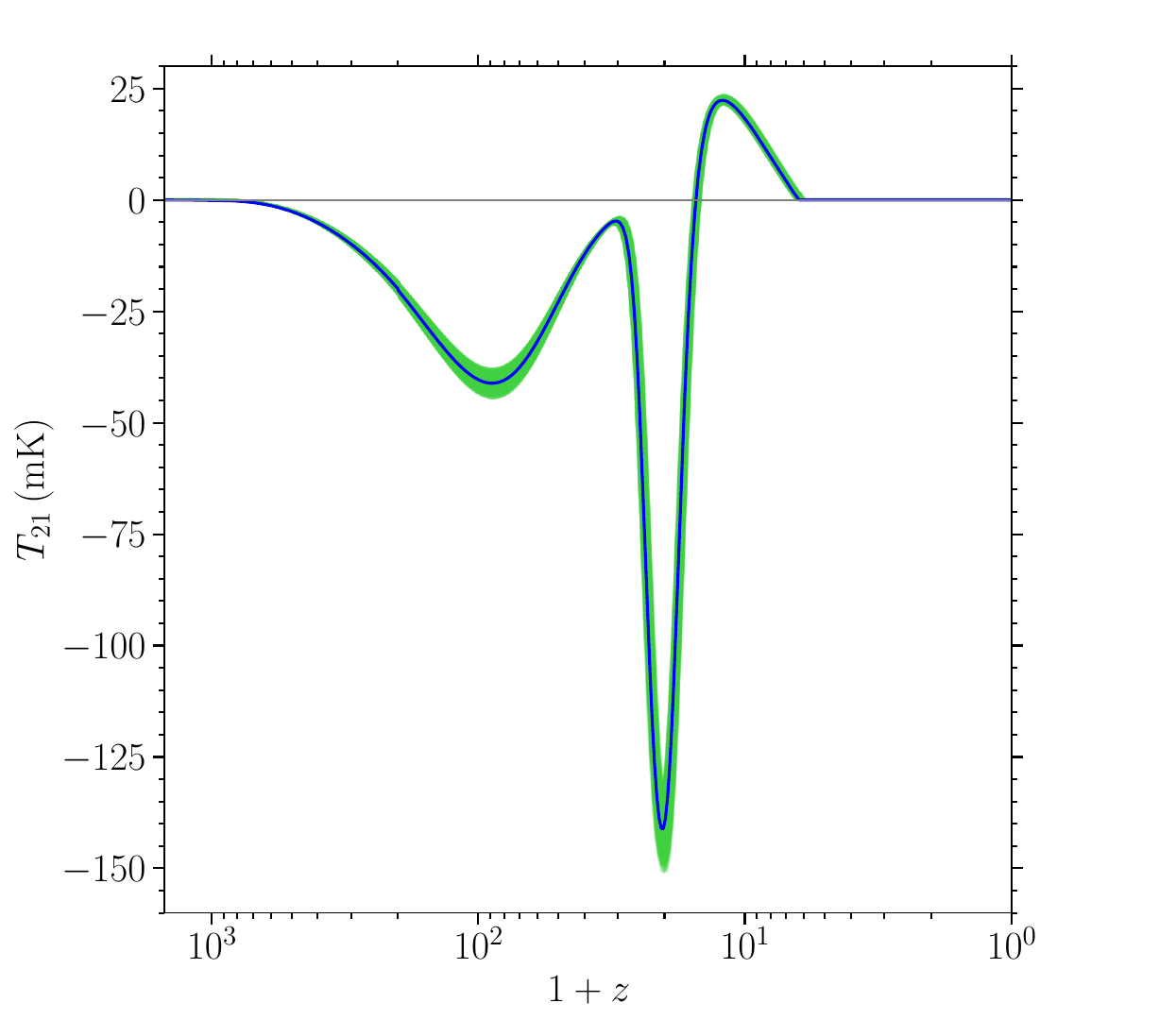}
\caption{2187 global signals (in green) generated with \texttt{ECHO21} for different cosmological parameters; we vary $H_0, \Omega_{\mathrm{m}}, \Omega_{\mathrm{b}}, \sigma_8, n_{\mathrm{s}}, T_{\gamma0}$, and $Y_{\mathrm{P}}$. See introductory section for the range of values we choose. The astrophysical parameters are fixed at their fiducial values. The solid-blue curve shows the fiducial signal (repeated from Fig.~\ref{fig:t21}).}\label{fig:cosmo}
\end{figure}

In \texttt{ECHO21} it is also possible to vary cosmological parameters just as easily. Similar to Fig.~\ref{fig:large}, where we showed signals for different astrophysical parameters, in Fig.~\ref{fig:cosmo} we show the signals for different cosmological parameters, namely the Hubble parameter ($H_0$), relative total matter density ($\Omega_{\mathrm{m}}$), relative baryon density ($\Omega_{\mathrm{b}}$), amplitude of density fluctuations ($\sigma_8$), spectral index of the primordial scalar spectrum ($n_{\mathrm{s}}$), CMB temperature today ($T_{\gamma0}$), and primordial helium fraction ($Y_{\mathrm{P}}$). We choose the range of parameters to be that set by the uncertainties (see the introduction section) with a total of three values for each parameter. Thus, if a parameter $p$ is $x\pm\epsilon$ then parameter values we explore are $\{x-\epsilon,x,x+\epsilon\}$. As a result, we have a total of $3^7=2187$ models. We run these models on 50 CPUs and found to complete in $\approx\SI{3}{\minute}$.

Because the cosmological parameters are so precisely determined (except for $H_0$), few works explore the impact of uncertainties in cosmological parameters on the 21-cm signal \citep[e.g.][]{Mirocha_2021}. As evident from Fig.~\ref{fig:cosmo}, the variation in the signal (corresponding to the uncertainties in \textit{Planck} data) around the fiducial signal model is small. Still, for the chosen uncertainties the strongest signal is $\SI{-151.9}{\milli\kelvin}$ (stronger by $\SI{9.6}{\milli\kelvin}$ than the fiducial model) while the weakest one is $\SI{-133.3}{\milli\kelvin}$ (weaker by $\SI{9.1}{\milli\kelvin}$ than the fiducial model) at cosmic dawn. In the dark ages, the strongest signal is $\SI{-44.40}{\milli\kelvin}$ and the weakest is $\SI{-37.93}{\milli\kelvin}$ both at $z\approx88$. Our findings are in close agreement with \citet{Mirocha_2021}.

Because of the experimental challenges associated with ground-based 21-cm experiments, there are prospects of setting up space- or Moon-based 21-cm experiments, e.g., \textit{DARE} \citep{Burns_2012}, \textit{PRATUSH} \citep{Rao_2023}, and \textit{CosmoCube} \citep{Artuc_2024}. Space-based 21-cm experiments can potentially dig deeper into the Universe providing a window beyond the cosmic dawn \citep{Koopmans_2021, Fialkov_2024}. Physics of the dark ages is expected to be much simpler making dark ages a clean testing ground for cosmological models. Since \texttt{ECHO21} allows one to explore the dark ages 21-cm signal for a space of cosmological parameters, it makes \texttt{ECHO21} an ideal tool to build synergies between 21-cm observations and other cosmological probes of the very early Universe such as BBN predictions, CMB anisotropies, polarization, or lensing \citep{Planck}. Relatedly, \texttt{ECHO21} can be used to synergise 21-cm cosmology with the well-known problem of the Hubble tension; local distance-ladder measurements lead to a higher measurement of the Hubble parameter ($H_0=74.03 \pm 1.42\,\mathrm{km s^{-1}Mpc^{-1}}$) which are in tension with the value inferred by \textit{Planck} collaboration by $\sim5\sigma$ \citep{Riess_2019, Abdalla_2022}.

\section{Conclusions}\label{sec:conc}
In this work we introduced a Python package called \texttt{ECHO21}. The approach taken by \texttt{ECHO21} is to simultaneously solve the differential equations governing the evolution of electron fraction and gas temperature since before the epoch of recombination until the present day accounting for the major processes that impact the histories. These include the standard recombination and photoionization terms throughout the timeline for the ionization history. Once star formation has commenced, we include the ionization due to UV and X-ray photons. For the thermal history, we account for the adiabatic cooling and Compton heating throughout the cosmic timeline. With the star formation in progress, we also account for X-ray heating and Ly$\upalpha$ heating. The latter is crucial in building an accurate 21-cm signal at cosmic dawn but has been ignored by many public codes. 

The main features of the code are as follows:
\begin{enumerate}
\item[1)] For a single set of astrophysical and cosmological parameters, the code outputs the CMB temperature, gas temperature, spin temperature, electron fraction, and the 21-cm signal throughout the dark ages, cosmic dawn, epoch of reionization until today. Further, it also outputs the volume-filling factor for the cosmic dawn period. For the \citet{PS74} choice of halo mass function (HMF), the total runtime is $\sim \mathcal{O}(1)\,$s.

\item[2)] In case of multiple parameters, the code outputs the corresponding global signals. As the code is (embarrassingly) MPI parallel, thousands of models can be generated in a few minutes provided an appropriate choice of high-performance computing resources.

\item[3)] We parametrize our model in terms of: the parameter to scale the number of Lyman series photons ($f_\mathrm{Ly}$), power-law index of Lyman series spectral energy distribution (SED, $s$), parameter to scale X-ray luminosity ($f_{\mathrm{X}}$), power-law index of X-ray SED ($w$), escape fraction of ionizing photons from galaxies ($f_{\mathrm{esc}}$), and a parameter which controls the star formation rate density (SFRD). See Table~\ref{tab:par}.

\item[4)] The code offers three choices of SFRD model, namely physically-motivated (default), semi-empirical and an empirically-motivated model. Further, for the physically-motivated and semi-empirical model, one has the option of choosing HMF such as Press-Schechter, Sheth-Tormen, Tinker, etc.

\item[5)] In addition to exploring 21-cm signals for a wide range of astrophysical parameters, \texttt{ECHO21} can be used to explore 21-cm signals for a wide range of cosmological parameters. These include $H_0, \Omega_{\mathrm{m}}, \Omega_{\mathrm{b}}, \sigma_8, n_{\mathrm{s}}, T_{\gamma0}$, and $Y_{\mathrm{P}}$, where the symbols have the usual meaning.
\end{enumerate}

The first version of \texttt{ECHO21} is based on a simple but sufficiently realistic\footnote{We have compared \texttt{ECHO21} with the semi-numerical code \href{https://www.cosmicdawnlab.com/21cmSPACE}{\texttt{21cmSPACE}} \citep{Fialkov_2012} and found reasonable agreement in key IGM quantities. For instance, using a semi-empirical SFRD model with press74 HMF, $f_{\mathrm{X}}=6.3$ and all other parameters remaining the same as in our fiducial model, the IGM gas temperature and the ionized fraction agree within approximately 13\% and 26\%, respectively. Furthermore, the redshift and depth of peak 21-cm absorption at CD agree to within 2\% and 20\%, respectively. We leave a more detailed comparative investigation for future work.} model of the intergalactic medium. The simplicity allows for a fast evaluation of global signal from the dark ages to the end of epoch of reionization. The current version of \texttt{ECHO21} is appropriate for making inferences and deriving constraints on astrophysical or cosmological parameters. (A primitive version of \texttt{ECHO21} has already been tested for parameter inference from \textit{EDGES} and \textit{SARAS}3 dataset.) It can be used to build synergies between 21-cm observations and probes of cosmological parameters or galaxy surveys. \texttt{ECHO21} is a developing code and future versions will see further improvements such as implementation of multiple scatterings of Ly$\upalpha$ photons, excess radio background model, option to work with more sophisticated star formation efficiency or escape fraction, among others.

\section*{Acknowledgements}
We thank Shikhar Asthana, Jiten Dhandha, Anastasia Fialkov, Prakash Gaikwad, Andrei Mesinger, Jordan Mirocha, and Sandro Tacchella for comments. It is also a pleasure to acknowledge discussions with several other members of the \textit{REACH} (Radio Experiment for the Analysis of Cosmic Hydrogen) Collaboration. SM is supported by the ERC (UKRI guaranteed) research grant EP/Y02916X/1. GK acknowledges support from the Kavli Institute Medium Term Visitor programme in Cambridge. GK is also partly supported by the Department of Atomic Energy (Government of India) research project with Project Identification Number RTI 4002. This work has been performed as part of the DAE-STFC collaboration `Building Indo-UK collaborations towards the Square Kilometre Array' (STFC grant reference ST/Y004191/1). This work used the DiRAC Data Intensive service (CSD3) at the University of Cambridge, managed by the University of Cambridge University Information Services on behalf of the STFC DiRAC HPC Facility (\href{https://dirac.ac.uk/}{www.dirac.ac.uk}). The DiRAC component of CSD3 at Cambridge was funded by BEIS, UKRI and STFC capital funding and STFC operations grants. DiRAC is part of the UKRI Digital Research Infrastructure.

\section*{Data Availability}
We release our Python package called \texttt{ECHO21}, available from PyPI and GitHub (\url{https://github.com/shikharmittal04/echo21.git}). The code documentation can be found at \url{https://echo21.readthedocs.io/en/latest/#}. 

\section*{Conflicts of Interest}
Authors declare no conflict of interest.

\bibliographystyle{mnras}
\bibliography{biblo} 

\appendix
\section{Ionization by X-ray photons}\label{app:xray_ion}
Computation of heating and ionization by the X-ray photons is expensive and slow to evaluate because of the multiple integrals involved \citep{Mesinger_2011, Mesinger_2013, Mittal_pbh}. As a result we need approximate empirical estimates which are quick to evaluate. This is often the case for X-ray heating \citep{F06, Mirocha_2015}, as we have used in this work (equation~\ref{eq:xheat}). However, a similar strategy for ionization rate ($\Gamma_\mathrm{X}$) has not been discussed in literature. We devise the following algorithm for a fast evaluation of $\Gamma_\mathrm{X}$. 

In physically-motivated approach, the standard X-ray heating is expressed as \citep{Madau_2017}
\begin{equation}
H_{\mathrm{X}}=4\pi\int_{E_0}^{E_1} (E-E_{\infty})\sigma(E)J_{\mathrm{X}}(E,z)\,\ud E\,,\label{Hx}
\end{equation}
in units of energy per unit time per neutral hydrogen atom. Similarly, the standard ionization (primary -- directly by the X-ray photons) rate is
\begin{equation}
\Gamma^{\mathrm{p}}_{\mathrm{X}}=4\pi\int_{E_0}^{E_1} \sigma(E)J_{\mathrm{X}}(E,z)\,\ud E\,,\label{Gx}
\end{equation}
where $E_{\infty}=\SI{13.6}{\electronvolt}$ is the ionization energy of hydrogen, $\sigma=\sigma(E)$ is the photoionization cross-section of \textsc{H\,i}--X-ray interaction, and $J_{\mathrm{X}}=J_{\mathrm{X}}(E,z)$ is the specific intensity (in terms of number of photons) of X-rays. The energy range of X-ray photons effective in heating and ionization of IGM is $E_0<E<E_1$. In this work we take $E_0$ and $E_1$ to be $0.2$ and $\SI{30}{\kilo\electronvolt}$, respectively.

In our chosen energy range the cross-section can be well-approximated by a power law, i.e., $\sigma\sim E^{u}$, where $u=-3.4$ \citep{Verner}. Similarly, if the SED (in terms of number) of X-ray photons emitted from star-forming galaxies is a power law of the form $\phi_\mathrm{X}\sim E^v$, where $v=-w-1$, then at any redshift specific intensity can be approximated to follow the same trend, so that $J_{\mathrm{X}}\sim E^v$. Thus, the ratio of heating and primary ionization rate is
\begin{equation}
\frac{H_{\mathrm{X}}}{\Gamma^{\mathrm{p}}_{\mathrm{X}}}=\frac{\int_{E_0}^{E_1} (E-E_{\infty})E^{u+v}\,\ud E}{\int_{E_0}^{E_1} E^{u+v}\,\ud E}\,.
\end{equation}
The above gives
\begin{equation}
E_w\equiv\frac{H_{\mathrm{X}}}{\Gamma^{\mathrm{p}}_{\mathrm{X}}}=\left(\frac{u+v+1}{u+v+2}\right)\left(\frac{E_1^{u+v+2}-E_0^{u+v+2}}{E_1^{u+v+1}-E_0^{u+v+1}}\right)-E_\infty\,.\label{eq:Ewgeneral}
\end{equation}
Inserting the values of $u$ and $v$ in the above we get $E_w$ as defined in equation~\eqref{eq:Ew}. Note that the ionization rate $\Gamma^{\mathrm{p}}_{\mathrm{X}}$ is enhanced by the secondary ionizations. These are enabled by the energetic electrons knocked out from neutral hydrogen by high-energy X-ray photons emitted from star-forming galaxies. The secondary ionization rate can be approximated as $f_{\mathrm{X,ion}}H_{\mathrm{X}}/E_\infty$. Accordingly, the total ionization rate is \citep{Kuhlen_2005}
\begin{equation}
\Gamma_{\mathrm{X}}\simeq\Gamma^{\mathrm{p}}_{\mathrm{X}}+\frac{f_{\mathrm{X,ion}}H_{\mathrm{X}}}{E_\infty}\,.\label{eq:gammax}
\end{equation}
Putting together equation~\eqref{eq:Ewgeneral} and \eqref{eq:gammax} we get
\begin{equation}
\Gamma_{\mathrm{X}}\simeq\left(\frac{1}{E_w}+\frac{f_{\mathrm{X,ion}}}{E_{\infty}}\right)\frac{q_{\mathrm{X}}}{f_{\mathrm{X,h}}n_{\ion{H}{i}}}\,,\label{eq:ionx}
\end{equation}
where $q_{\mathrm{X}}=f_{\mathrm{X,h}}n_{\ion{H}{i}}H_{\mathrm{X}}$ is the effective volumetric heating rate, for which we already have an estimate based on empirical formalism (see equation~\ref{eq:xheat}). Thus, having computed $q_{\mathrm{X}}$ it is trivial to obtain $\Gamma_{\mathrm{X}}$ via equation~\eqref{eq:ionx}.

\bsp	
\label{lastpage}
\end{document}